\documentclass[12pt]{iopart}

\usepackage{iopams}

\usepackage{xcolor}
\usepackage{tikz}
\usepackage{verbatim}
\usetikzlibrary{arrows,trees,decorations.markings,patterns}
\usepackage{caption}

\begin{document}

\paper[Consistency and Causality of Interconnected Nonsignaling Resources]{Consistency and Causality of Interconnected Nonsignaling Resources}

\author{Peter Bierhorst}

\address{Mathematics Department, University of New Orleans, 2000 Lakeshore Drive, New Orleans, LA 70148 }
\ead{plbierho@uno.edu}
\vspace{10pt}

\begin{abstract}
This paper examines networks of $n$ measuring parties sharing $m$ nonsignaling resources that can be locally wired together: that is, each party follows a scheme to measure the resources in a cascaded fashion with inputs to later resources possibly depending on outputs of earlier-measured ones. A specific framework is provided for studying probability distributions arising in such networks, and this framework is used to directly prove some accepted, but often only implicitly invoked, facts: there is a uniquely determined and well-defined joint probability distribution for the outputs of all resources shared by the parties, and this joint distribution is nonsignaling. It is furthermore shown that is often sufficient to restrict consideration to only extremal nonsignaling resources when considering features and properties of such networks. Finally, the framework illustrates how the physical theory of nonsignaling boxes and local wirings is \textit{causal}, supporting the applicability of the inflation technique to constrain such models. 

For an application, we probe the example of (3,2,2) inequalities that witness genuine three-party nonlocality according to the local-operations-shared-randomness definition, and show how all other examples can be derived from that of Mao \textit{et al}.~(2022 \textit{Phys.~Rev.~Lett.}~\textbf{129} 150401). 
\end{abstract}

%
%
\submitto{\jpa}
%
%
%

\section{Introduction: nonsignaling resources and networks}

Quantum mechanics is \textit{nonlocal} in the sense that certain quantum experiments involving spatially separated measuring parties do not admit a local hidden variable description \cite{BELL,CHSH}. Quantum nonlocality experiments have been performed under strict conditions \cite{hensen:2015,shalm:2015,giustina:2015,rosenfeld:2017}, confirming the phenomenon.

Quantum nonlocality experiments, while demonstrating strange correlations between distant parties, still satisfy a condition known as the \textit{nonsignaling} condition: a measuring party cannot exploit a quantum experiment to send signals to a spatially separated party. This assures compliance with special relativity in experiments where separated measurements can be performed near-simultaneously at great distances such that any signals would need to be traveling faster than the speed of light. The study of the non-signaling condition is of interest in various contexts, such as abstract definitions of nonlocality not invoking quantum mechanics \cite{barrettpironio:2005}, attempts to derive quantum mechanics from physical principles \cite{PRBOX}, and certifying unpredictability of random numbers under minimal assumptions \cite{bierhorst:2018}.

The nonsignaling condition can be stated formally as follows: consider an experiment of two spatially separated parties named Alice and Bob, each of whom performs a measurement using an apparatus; each apparatus has a choice of measurement setting labeled (respectively) $X$ and $Y$ and provides a measurement outcome labeled (respectively) $A$ and $B$. Then the joint probability distribution $\mathbb P(A,B|X,Y)$ of outcomes conditioned on settings is \textit{nonsignaling} if each party's marginal outcome probabilities are independent of settings choices of the other. Mathematically this can be expressed as the equalities
\begin{eqnarray}\label{e:nosig2}
\mathbb P(A=a|X=x,Y=y) &=& \mathbb P(A=a|X=x,Y=y')\nonumber\\
\mathbb P(B=b|X=x,Y=y) &=& \mathbb P(B=b|X=x',Y=y)
\end{eqnarray}
which hold for all values $a,x,y$ and $x',y'$, where a marginal probability such as ${\mathbb P(A=a|X=x,Y=y)}$ is obtained from the joint distribution by the summation ${\sum_b\mathbb P(A=a,B=b|X=x,Y=y)}$. All distributions $\mathbb P$ obtainable with quantum mechanics satisfy the nonsignaling condition \eref{e:nosig2}, but the converse is not true: there are distributions satisfying \eref{e:nosig2} that cannot be observed through measurements of entangled quantum states, such as the paradigmatic example of the Popescu-Rohrlich ``PR box" distribution \cite{PRBOX}. \eref{e:nosig2} can be generalized to scenarios of $n>2$ separated measuring parties, whereby it is stipulated that the outcome distribution of any given subset of the $n$ parties is required to be independent of the settings of the other parties.

An interesting question is what sort of probability distributions can be observed in a three-party experiment for which each pair of parties share bipartite nonsignaling resources satisfying \eref{e:nosig2} -- possibly multiple such resources, allowing local ``wirings" whereby each party can access their resources in cascaded fashion and condition inputs provided to later resources on observed outputs from earlier ones. Early results on this question can be found in Section IIIC of \cite{barrett:2005} and \cite{barrettpironio:2005}. Recently, the question is of renewed interest in light of arguments \cite{bierhorst:2021,coiteux:2021} that only three-party probability distributions that \textit{cannot} be replicated by such underlying networks of bipartite resources -- possibly supplemented with global shared randomness -- should be considered \textit{genuinely multipartite nonlocal} (GMNL). This approach resolves an anomaly \cite{coiteux:2021} in earlier definitions of the GMNL concept \cite{svetlichny:1987,seevink:2002,bancal:2013} in which parallel independent two-party nonlocality experiments can be counterintuitively classified as GMNL. The new revised notion of GMNL is named in \cite{coiteux:2021a} as LOSR-GMNL, with LOSR standing for local operations and shared randomness; quantum measurements of the three-way entangled GHZ state \cite{GHZ} can exhibit LOSR-GMNL \cite{chao:2017,bierhorst:2021,coiteux:2021} and recent experiments \cite{mao:2022,cao:2022,huang:2022} provide some initial evidence of the phenomenon.  

Motivated in part by the LOSR-GMNL definition, this paper studies the general question of how to systematically model $n$-party conditional distributions, or \textit{behaviors}, of the form $\mathbb P(\mathbf A_1, ..., \mathbf A_n|\mathbf X_1,..., \mathbf X_n)$ that are induced as follows: a network of $m$ nonsignaling resources, each shared by a subset of the parties, is measured by the parties in cascaded fashion after each party $i$ receives a setting $\mathbf X_i$; then, each party's final outcome $\mathbf A_i$ is a function of the observed outputs from the resources. This scenario can be referred to as \textit{nonsignaling resources with local wirings}. The study of behaviors obtained this way essentially reduces to the study of the joint distribution of all the resource outputs in the underlying network: $\mathbb P(\vec A_1, .... ,\vec A_m|\mathbf X_1,..., \mathbf X_n)$ where $\vec A_j$ refers to all the outputs of the $j$th shared resource.

The central goal of this work is thus to provide a framework for direct study of the joint probability distributions $\mathbb P(\vec A_1, .... ,\vec A_m|\mathbf X_1,..., \mathbf X_n)$ so as to 1) directly prove here some commonly accepted properties of such distributions, and 2) provide a foundation for rigorous future results about them. Regarding point (1), it has been accepted, often tacitly, that there \textit{is} a well-posed joint distribution $\mathbb P(\vec A_1, .... ,\vec A_m|\mathbf X_1,..., \mathbf X_n)$ which is itself nonsignaling and causal (causality here roughly corresponding to an intuitive notion that the marginal distribution of a subset of parties is determined only by the resources they measure and how they measure them; quantum mechanics is an example of a causal theory but other more exotic theories may be causal as well). 
But considering that for wired \textit{signaling} resources, a consistent joint distribution is not in general possible (see Figure 1 of Ref.~\cite{bancal:2013} for an example), it is good to be clear about why this is true when the resources are nonsignaling. Accordingly, this work rigorously demonstrates that the joint distribution induced by a network of wired nonsignaling resources is well-defined and itself nonsignaling. Moreover, while previous works such as \cite{janotta:2012,chiribella:2011,chiribella:2014} are cited in \cite{coiteux:2021,coiteux:2021a} to justify the \textit{causality} of the paradigm, these previous works are somewhat abstract and do not always address the point directly. The framework of this paper provides a clear foundation for demonstrating the causality of the theory of wired nonsignaling resources. An important consequence of the causality of the paradigm is that it enables use of the powerful inflation technique \cite{wolfe:2019}, which applies to causal theories.

Indeed, while this work reinforces the fact that that constraints derived from the inflation technique \cite{wolfe:2019} are valid in constraining behaviors in networks of wired nonsignaling resources, it will provide an important foundation for deriving constraints satisfied by \textit{only} these behaviors but possibly violated by different causal theories -- an important example being scenarios allowing for entangled measurements and/or generalizations thereof. Since the inflation technique applies to all causal theories, it cannot readily be used to address this separation. Thus the framework introduced here for direct study of \textit{just} wired nonsignaling resources will be useful in resolving the question of when/whether behaviors that can be observed in generalized probabilistic theories \textit{with} entangled measurements (or generalized analogues thereof) can \textit{not} be observed in networks where these are prohibited (such as nonsignaling resources with local wirings). This corresponds to the question of whether there are behaviors in regions $\mathcal R_3/\mathcal R_5$ in the Venn diagram of Figure 2 of \cite{bierhorst:2023}; Conjecture 1 in Section V-C of \cite{coiteux:2021a} is an argument that $\mathcal R_5$ is nonempty. The question is somewhat subtle as some behaviors that would seemingly require entangled measurements -- such as the device-independent certification of entangled measurements protocol of Ref.~\cite{rabelo:2011} -- can be counterintuitively simulated with wired nonsignaling boxes \cite{bierhorst:2023}. Study of this region will increase our understanding of entangled measurements; it is also motivated by a variant definition \cite{bierhorst:2021} of LOSR-GMNL in which entangled measurements and generalizations thereof are not allowed for the class of behaviors that are classified as bipartite-only nonlocal.

Note that since quantum resources are nonsignaling, any constraint proved in this context will apply to a practical scenario of networks of quantum-achievable nonsignaling resources measured in cascaded fashion -- the ``quantum box'' paradigm of the set $QB_2$ in \cite{bierhorst:2023}, which is directly relevant to the proposed definition of genuine network nonlocality given in \cite{supic:2022}.

The paper starts by defining nonsignaling resources and networks thereof in Section \ref{s:defandconsist}, where a method for determining the joint distribution is formalized and shown to be consistent. Section \ref{s:netproperties} derives properties of joint distribution: the nonsignaling property, the ability to restrict attention to extremal nonlocal nonsignaling resources in certain cases (an important technique that was used in, for example, \cite{bierhorst:2021}), and a discussion of the causality of the framework which supports the applicability of the inflation technique.

The paper concludes with a case study example: inequalities witnessing LOSR-GMNL in the three-party scenario. It is shown that all known three-party inequalities with two settings per party and two outcomes per setting -- the simplest possible scenario witnessing LOSR-GMNL (see Section SM 3 of \cite{bierhorst:2023}) -- can be derived from the inequality of of Mao \textit{et al.}~\cite{mao:2022} (which was obtained with the inflation technique, and so the causality results of this paper reinforce the applicability of this inequality to the paradigm of wired nonsignaling resources). These derivable inequalities include the inequality of Chao and Reichardt \cite{chao:2017} as formulated in \cite{bierhorst:2021} (which is notable as the Chao-Reichardt inequality had previously only been derived directly within the paradigm of wired nonsignaling resources; by deriving it here as a consequence of the inequality of Mao \textit{et al.} we show it holds of the more broad class of causal theories), and inequality (1) of Cao \textit{et al.}~\cite{cao:2022}. A second inequality of Cao \textit{et al.}, which has an extra setting for one of the parties, can also be derived from that of Mao \textit{et al.}; a natural open question is whether different inequalities can be discovered in this scenario.

\section{Nonsignaling resources: definition, a framework for studying networks, and consistency of the joint distribution} \label{s:defandconsist}

We are interested in behaviors $\mathbb P(\mathbf{A}_1, ..., \mathbf{A}_n |\mathbf{X}_1, ..., \mathbf{X}_n)$ that can be induced by underlying networks of nonsignaling resources. We will notate the distributions of the underlying network resources with $R$, as in $R(ABC|XYZ)$, to distinguish these from the final global distribution $\mathbb P$. It is also helpful to refer to the variables of the resource occurring in $R(\cdots|\cdots)$ as \textit{outputs} and \textit{inputs}, to distinguish them from the variables of the overall behavior $\mathbb P(\mathbf{A}_1, ..., \mathbf{A}_n |\mathbf{X}_1, ..., \mathbf{X}_n)$, for which we call $\mathbf{X_i}$ the \textit{setting} and $\mathbf A_i$ the \textit{outcome} or \textit{final outcome}. In the next subsection, we introduce a formal definition of nonsignaling for an $n$ party resource $R$ that generalizes \eref{e:nosig2}, and derive some important consequences of the nonsignaling condition.

\subsection{Properties of nonsignaling resources}\label{s:properties}

For an $n$-party resource $R(\cdots|\cdots)$, the nonsignaling condition is as follows: for each $j$ in $\{1,...,n\}$ and each pair of possible values $x_j$ and $x'_j$ that the input choice $X_j$ can assume, we have
\begin{equation*}
\fl\quad\sum_{a_j}R(A_1=a_1, ...,A_j=a_j,..., A_n=a_n|X_1=x_1,...,\underbrace{X_j=x_j}_\textnormal{change},..., X_n=x_n)
\end{equation*}
\begin{equation}\label{e:nosig}
\fl=\sum_{a_j}R(A_1=a_1, ...,A_j=a_j,..., A_n=a_n|X_1=x_1,...,\overbrace{X_j=x'_j},..., X_n=x_n)  
\end{equation}
for each fixed choice of $x_i$ among $i\ne j$. In words, this means that the conditional distribution of the $A_i$ \textit{excluding} $A_j$ is independent of party $j$'s input choice. This represents the idea that one party (the $j$th) cannot signal to the rest through their choice of input.  For the rest of the paper, we will use a shorthand in expressions like \eref{e:nosig} whereby $R(\vec a|\vec x)=R(a_1, ...,a_n|x_1,..., x_n)$ is shorthand for $R(A_1=a_1, ..., A_n=a_n|X_1=x_1,..., X_n)$.

A few points are worth mentioning before moving on. First, use of the conditional distribution notation $R(\vec a|\vec x)$ suggests the existence of a joint distribution of all random variables comprising $\vec A$ and $\vec X$ from which the conditional probabilities are derived. However, in studies of networked nonsignaling resources, a full probability distribution of the inputs $\vec X$ is somewhat besides the point (even if it may exist) -- we want to think of the inputs more as choices one can supply to the resources to which they respond. In this sense $R(\cdots|\cdots)$ is perhaps better thought of as a family of (unconditional) probability distributions of random variables $\vec A$, merely indexed by $\vec x$. Thus we avoid tacitly appealing to input probability distributions in the derivations below, so that consequences of \eref{e:nosig} below could just as easily apply to $\vec x_n$-indexed families of (unconditional) probability distributions ``$R_{\vec x_n}(\vec A_n)$" satisfying a suitably re-notated \eref{e:nosig}, while keeping the standard convention of conditional probability notation $R(\cdots|\cdots)$.

\begin{figure}
\captionsetup{singlelinecheck = false, justification=justified}

\begin{tikzpicture}[scale=1]

\def\sigdist{15}
\def\eqtri{1.1}
 
\def\ax{-1*\eqtri}    
\def\ay{0}
   
\def\bx{-.3090*\eqtri}    
\def\by{.9511*\eqtri}   
   
\def\cx{.8090*\eqtri}    
\def\cy{.5878*\eqtri}

\def\dx{.8090*\eqtri}    
\def\dy{-.5878*\eqtri}
   
\def\ex{-.3090*\eqtri}    
\def\ey{-.9511*\eqtri}

\node[circle,draw=black,minimum size=1mm] at (\ax,\ay) (a1) {};
\node at (\ax,\ay) {\tiny$A1$};
\node[circle,draw=black,minimum size=1mm] at (\bx,\by) (a2) {};
\node at (\bx,\by) {\tiny$A2$};
\node[circle,draw=black,minimum size=1mm] at (\cx,\cy) (a3) {};
\node at (\cx,\cy) {\tiny$A3$};
\node[circle,draw=black,minimum size=1mm] at (\dx,\dy) (a4) {};
\node at (\dx,\dy) {\tiny$A4$};
\node[circle,draw=black,minimum size=1mm] at (\ex,\ey) (a5) {};
\node at (\ex,\ey) {\tiny$A5$};

\node[rectangle,rounded corners,draw] at (-\sigdist*.4,0) (source) {\tiny Source} ;
\node[rectangle,rounded corners,draw] at (\sigdist*.6,0) (receiver) {\tiny Receiver};

\def\bep{\sigdist*15}   
\def\ber{\sigdist*.3} 

  \draw[shorten >=\bep,shorten <=\ber,->,dashed] (a2) -- (receiver);
  \draw[shorten >=\bep,shorten <=\ber,->,dashed] (a5) -- (receiver);
  
\def\cdp{\sigdist*12.5}   
\def\cdr{\sigdist*.3} 
  
  \draw[shorten >=\cdp,shorten <=\cdr,->,dashed] (a3) -- (receiver);
  \draw[shorten >=\cdp,shorten <=\cdr,->,dashed] (a4) -- (receiver);
  
  \def\gapdist{-1.5}
  \draw[|-|,thick,color=red] (\ax,\gapdist)-- node[below]{\tiny \bf Gap} (\bx,\gapdist);

  \def\lbep{\sigdist*2.5 }   
\def\lber{\sigdist*13} 

  \draw[shorten >=\lbep,shorten <=\lber,->,dashed] (a2) -- (receiver);
  \draw[shorten >=\lbep,shorten <=\lber,->,dashed] (a5) -- (receiver);
  
\def\lcdp{\sigdist*.5}   
\def\lcdr{\sigdist*12.5} 
  
  \draw[shorten >=\lcdp,shorten <=\lcdr,->,dashed] (a3) -- (receiver);
  \draw[shorten >=\lcdp,shorten <=\lcdr,->,dashed] (a4) -- (receiver);

    \draw[->,dashed]  (-\sigdist*.35,0)-- node[font={\tiny},above=5mm,align=left]{(1) Source\\ emits a signal}  (-\sigdist*.31,0);
      \draw[->,dashed]  (-\sigdist*.15,0)-- (-\sigdist*.11,0);

    \node[font={\tiny},align=left] at (-1.8*\eqtri,.9*\eqtri) {(2) $A1$ uses signal \\ to determine \\ input choice $X_1$};
    \node[font={\tiny},align=left] at (2.6*\eqtri,1*\eqtri) {(3)$A2$-$A5$ immediately \\ transmit outputs $A_2$-$A_5$};
    \node[font={\tiny},align=left] at (7*\eqtri,.9*\eqtri) {(4) Receiver reconstructs \\ information about signal \\using $A_2$-$A_5$};

  \end{tikzpicture}

\caption{\textbf{Faster-than-light signaling through violation of \eref{e:nosig}.} In the scheme above, the signal distance between Source and Receiver is effectively shortened by the length of the span marked ``Gap," with a small $\epsilon$ correction owing to slight diagonality in the paths. The correction vanishes asymptotically as the distance to Receiver increases. If all dashed-line signals travel at the speed of light, signal information traverses the Source-Receiver span faster than the speed of light. Sources and receivers positioned along different diagonal axes motivate \eref{e:nosig} for other choices of the signaling input $X_j$ ($j=1$ above). A similar figure with a regular $n$-gon motivates \eref{e:nosig} for a set of $n$ parties.}\label{f:pentagonsignal}
\end{figure}
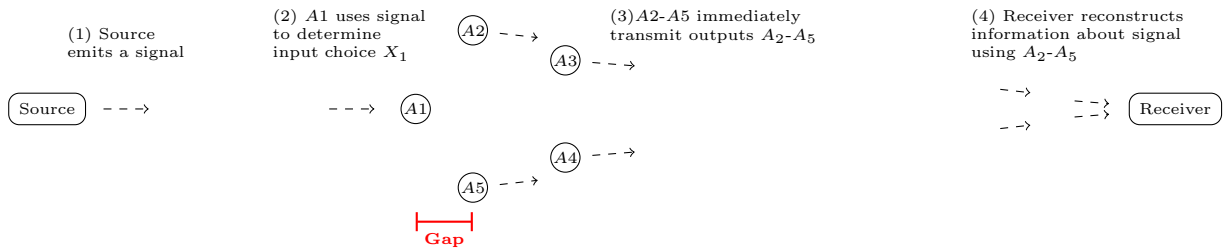

It also merits briefly discussing the physical motivation of \eref{e:nosig}. The justification via special relativity can be seen by considering a scenario where the $n$ parties are arranged at the vertices of a regular polygon, in which case any violation of \eref{e:nosig} could result in a signal-speed boost in a particular direction: see Figure \ref{f:pentagonsignal} for an illustration in the case of five parties. The figure may make \eref{e:nosig} seem incomplete, as there are of course other signalings that could be well motivated, such as many-to-one (running the figure in reverse), or a group of two adjacent parties signaling to the remaining three; conversely, certain other subset-to-subset signaling prohibitions might not be so clear how to intuitively justify based solely on special relativity considerations in the context of Figure \ref{f:pentagonsignal}. However, it is known (see \cite{barrett:2005} Section IIIA) that the other subset-to-subset signaling prohibitions can be derived mathematically from \eref{e:nosig}, and so once the \eref{e:nosig} condition is accepted, one does not require new physical motivation to accept other nonsignaling conditions.

We now derive important consequences of \eref{e:nosig}. First, a more general prohibition on subset-to-subset signaling among the parties can be formulated in the following manner: For $1\le p\le n$, let $\vec a_p$ denote $a_1,...,a_p$ and $\vec a_q$ denote $a_{p+1},...,a_n$, and let $\vec x_p$ and $\vec x_q$ denote the corresponding sets of $x_i$ variables. Then for any fixed choice of $\vec a_p, \vec x_p$, $\vec x_q$, and $\vec x'_q \ne \vec x_q$, we can prove
\begin{equation}\label{e:nosigsubcomp}
\sum_{\vec a_q} R(\vec a_p, \vec a_q | \vec x_p, \vec x_q)=\sum_{\vec a_q} R(\vec a_p, \vec a_q | \vec x_p, \vec x'_q).
\end{equation}
The condition above applies to any partition of the parties into two sets. The proof of \eref{e:nosigsubcomp}, which we write out in \ref{s:ap1}, amounts to iterated applications of \eref{e:nosig}. Condition \eref{e:nosigsubcomp} can be equivalently re-written a little more compactly in terms of the marginal distribution of $\vec a_p$ as 
\begin{equation}\label{e:compcomp}
R(\vec a_p| \vec x_p, \vec x_q)=R(\vec a_p | \vec x_p, \vec x'_q).
\end{equation}
With this in mind, we can define the probability distribution $R(\vec a_p|\vec x_p)$ as
\begin{eqnarray}\label{e:defreduced}
R(\vec a_p|\vec x_p) &:=& R(\vec a_p| \vec x_p, \vec x_q)
\end{eqnarray}
for some fixed choice of $\vec x_q$ -- any choice of which will do, as \eref{e:compcomp} ensures there is no ambiguity in leaving this choice arbitrary. We note that if a distribution over $\vec X$ is presumed, so that a joint distribution of all random variables $\vec A, \vec X$ exists and the conditional probabilities $R(\vec a_p|\vec x_p)$ can be calculated directly from it, satisfaction of \eref{e:nosig} ensures that these calculations will be consistent with \eref{e:defreduced}; see \ref{s:ap1} for a demonstration. Finally, and unsurprisingly, the reduced distribution \eref{e:defreduced} is itself nonsignaling: Subdividing $\vec a_p$ into two strings $\vec a_{p_r}$ (receiver) and $\vec a_{p_s}$ (signaler), we have:
\begin{eqnarray}
\sum_{\vec a_{p_s}}R(\vec a_{p_r},\vec a_{p_s}|\vec x_{p_r},\vec x_{p_s})  &=&R(\vec a_{p_r}|\vec x_{p_r},\vec x_{p_s}) \nonumber\\ 
&=& R(\vec a_{p_r}|\vec x_{p_r},\vec x_{p_s},\vec x_q) \nonumber\\ 
&=& R(\vec a_{p_r}|\vec x_{p_r},\vec x'_{p_s},\vec x'_q) \nonumber\\ 
&=& R(\vec a_{p_r}|\vec x_{p_r},\vec x'_{p_s}) \qquad=\sum_{\vec a_{p_s}}R(\vec a_{p_r},\vec a_{p_s}|\vec x_{p_r},\vec x'_{p_s}) \label{e:subtosub}
\end{eqnarray}
where we applied \eref{e:defreduced}, then \eref{e:compcomp}, then \eref{e:defreduced} after converting the sums into equivalent expressions about marginal probabilities. The equality $R(\vec a_{p_r}|\vec x_{p_r},\vec x_{p_s}) = R(\vec a_{p_r}|\vec x_{p_r},\vec x'_{p_s})$ can be given an operational interpretation as the broadest notion of nonsignaling, encapsulating the idea that no subset of parties (those corresponding to $\vec a_{p_s}$) can signal to any disjoint other subset (those corresponding to $\vec a_{p_r}$) without explicit reference to the third uninvolved subset of remaining parties (those corresponding to $\vec a_q$).

The final property that we derive, required for some arguments in the next section, is as follows: the input-conditional distribution of a subset of parties, conditioned additionally on the inputs \textit{and outputs} of the other parties, is nonsignaling. That is, assume a particular set of outputs $\vec a_q$ of the last $q$ parties occurs with nonzero probability, given the input vector $\vec x_q$: $R(\vec a_q|\vec x_q)>0$. Then the natural definition of the distribution of the first $p$ parties' output $\vec a_p$ conditioned on their input $\vec x_p$, as well as the inputs $\vec x_q$ and outputs $\vec a_q$ of the $q$ group, is
\begin{equation}\label{e:condsetout}
R^{\vec a_q, \vec x_q} (\vec a_p| \vec x_p)=R(\vec a_p|\vec x_p, \vec x_q, \vec a_q):=\frac{R(\vec a_p,\vec a_q| \vec x_p,\vec x_q)}{R(\vec a_q|\vec x_q)}
\end{equation}
where we exploit \eref{e:defreduced} to justify writing $R(\vec a_q|\vec x_q)$ instead of the more generally valid $R(\vec a_q|\vec x_p, \vec x_q)$ in the denominator above. It is immediate that $R^{\vec a_q, \vec x_q} (\vec a_p| \vec x_p)$ is a valid probability distribution (nonnegative and sums to one over $\vec a_p$). Furthermore $R^{\vec a_q, \vec x_q} (\vec a_p| \vec x_p)$ is no-signaling as follows:
\begin{eqnarray}
\sum_{a_p}R(\vec a_{p-1},a_p| \vec x_{p-1},x_p, \vec a_q, \vec x_q)
&=&\frac{\sum_{a_p}R(\vec a_{p-1},a_p, \vec a_q|\vec x_{p-1},x_p, \vec x_q)}{R(\vec a_q| \vec x_q)}\nonumber\\
&=&\frac{\sum_{a_p}R(\vec a_{p-1},a_p, \vec a_q|\vec x_{p-1},x'_p, \vec x_q)}{R(\vec a_q| \vec x_q)}\nonumber\\
&=&\sum_{a_p}R(\vec a_{p-1},a_p| \vec x_{p-1},x'_p, \vec a_q, \vec x_q)\label{e:outcomeconditional}
\end{eqnarray}
where in the middle equality we apply \eref{e:nosig} to the numerator. As \eref{e:outcomeconditional} is equivalent to \eref{e:nosig} when applied to the reduced distribution $R^{\vec a_q, \vec x_q} (\vec a_p| \vec x_p)$, it naturally follows that $R^{\vec a_q, \vec x_q} (\vec a_p| \vec x_p)$ satisfies all of the properties \eref{e:nosigsubcomp} -- \eref{e:subtosub} derived as a consequence of \eref{e:nosig}. Furthermore it is straightforward to confirm that conditioning as in \eref{e:condsetout} iteratively is equivalent to performing the steps all at once; i.e., if $\vec a = (\vec a_p,\vec a_q, \vec a_r)$ and we take the no-signaling distribution $R^{\vec a_r, \vec x_r} (\vec a_p, \vec a_q| \vec x_p, \vec x_q)$ and condition on the input-output combination $\vec a_q,\vec x_q$ to get $(R^{\vec a_r, \vec x_r})^{\vec a_q,\vec x_q} (\vec a_p| \vec x_p)$, the result is equivalent to $R^{\vec a_q\vec a_r, \vec x_q \vec x_r} (\vec a_p| \vec x_p)$.

\subsection{Networked collections of nonsignaling resources: paths and decision trees}\label{s:decisiontree}

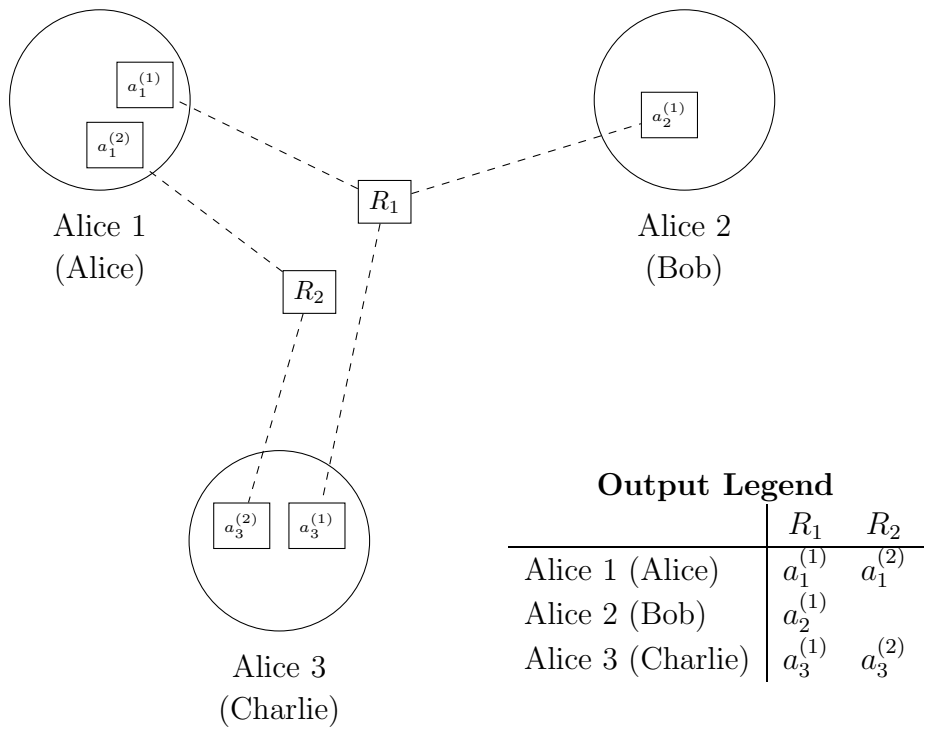
\begin{figure}
    \centering

\begin{tikzpicture}[scale=.2]

\def\radi{6}
\def\eqtri{17} 
\def\ax{-.866*\eqtri*1.15}    
\def\ay{.5*\eqtri*1.15}
   
\def\bx{.866*\eqtri*1.15+5}    
\def\by{.5*\eqtri*1.15}
   
\def\cx{-5}    
\def\cy{-\eqtri*1.15}   

\node[draw] at (2,3) (src1) {\footnotesize$R_1$};
\node[draw] at (-3,-3) (src2) {\footnotesize $R_2$};
\node[draw] at (\ax+1,\ay-3) (a2) {\tiny$a^{(2)}_1$};
\node[draw] at (\cx-2.5,\cy+1) (c2) {\tiny$a^{(2)}_3$};
\node[draw] at (\ax+3,\ay+1) (a1) {\tiny$a^{(1)}_1$};
\node[draw] at (\bx-1,\by-1) (b1) {\tiny$a^{(1)}_2$};
\node[draw] at (\cx+2.5,\cy+1) (c1) {\tiny$a^{(1)}_3$};

\node at (24,-22)[align=center]{\textbf{Output Legend}\\\begin{tabular}{l|cc}
& $R_1$ & $R_2$\\
\hline
Alice 1 (Alice)& $a^{(1)}_1$ &  $a^{(2)}_1$ \\
Alice 2 (Bob)& $a^{(1)}_2$ &   \\
Alice 3 (Charlie)& $a^{(1)}_3$ &$a^{(2)}_3$  \\
\end{tabular}};

\draw[dashed] (src1) -- (a1);
\draw[dashed] (src1) -- (b1);
\draw[dashed] (src1) -- (c1);
\draw[dashed] (src2) -- (a2);
\draw[dashed] (src2) -- (c2);

  \draw[fill=none](\ax,\ay) circle (\radi) node [black,yshift=-2cm,align=center] {Alice 1\\ (Alice)};
  \draw[fill=none](\bx,\by) circle (\radi) node [black,yshift=-2cm,align=center] {Alice 2\\(Bob)};
  \draw[fill=none](\cx,\cy) circle (\radi) node [black,yshift=-2cm,align=center] {Alice 3\\(Charlie)};
  
  \end{tikzpicture}
  
\caption{\textbf{An example of three parties sharing two nonsignaling resources.}}\label{f:3wayexample}
\end{figure}

Having defined the nonsignaling condition \eref{e:nosig} and derived some of its consequences, we are ready to study networked collections of nonsignaling resources. We consider a scenario of $n$ spatially separated measuring parties, which we call Alice 1 through Alice $n$ (or Alice-Bob-Charlie in scenarios of $n=3$ parties). The $n$ parties share a set of $m$ nonsignaling resources $\mathcal R = \{R_1, ..., R_m\}$; Figure \ref{f:3wayexample} gives a schematic example of $n=3$ parties sharing $m=2$ resources. Each nonsignaling resource $R_k$ is shared by a subset of parties indexed by a set $\mathcal M_k = \{k_1,...,k_{n_k}\} \subseteq \{1, ...,n\}$ whose cardinality $n_k$ can be as small as 1 and as large as $n$; in the example Figure \ref{f:3wayexample} these sets correspond to columns in the ``Output Legend" of Fig.~\ref{f:3wayexample}, so the set $\mathcal M_1$ is all three parties while $\mathcal M_2$ is only Alice and Charlie. Each party sharing the resource $R_k$ has an \textit{input} $X^{(k)}_{k_j}$ that can take one or more more values $x^{(k)}_{k_j}$, for which there is then a corresponding \textit{output} $A^{(k)}_{k_j}$ taking values $a^{(k)}_{k_j}$. Below, we will omit the superscripts $(k)$ from the $X$ and $A$ variables when it is clear from context to which $R_k$ they are associated. We assume that the output space for a fixed $A^{(k)}_p$ is the same for every choice of input $x^{(k)}_p$. (This is not restrictive, because we can always make it true of a resource by artificially augmenting value spaces, assigning probability zero to the added outputs; \eref{e:nosig} will hold of the augmented distribution.)

We want to examine the joint probability distributions that arise when each party can measure the portion of the resources they share in different orders and use outputs of earlier resources to influence choices of inputs provided to later resources, as well as the order in which they access the later resources. Mathematically: letting $\vec A_k$ denote the vector of all outputs of resource $k$ possessed by the subset of parties $\mathcal M_k$, we are interested in the distribution of joint outputs of all $n$ resources $\mathbb P(\vec A_1, ..., \vec A_m|\mathbf X_1, ..., \mathbf X_n)$ when each party follows such a scheme. A fixed party $p$ observes one entry from each resource-output vector $\vec A_k$ that corresponds to a resource $R_k$ they share; $\mathbf X_p$ denotes an initial setting provided to the $p$th party, on which they can condition their strategy for accessing the resources. Recall that we call $\mathbf X_p$ the \textit{setting} to differentiate it from the various $X^{(k)}_p$ supplied to the $R_k$ resources which we call \textit{inputs}. 

We model the scheme with a decision tree for each party $p$, which encompasses their strategy for accessing their resources: which resource $R_k$ she will access first, then how she will proceed to the next resource depending on the observed output, and so forth. After carefully stipulating the structure of these decision trees, we provide a formula for computing $\mathbb P(\vec A_1, ..., \vec A_m|\mathbf X_1, ..., \mathbf X_n)$, and show that the procedure is sound (i.e., leads to a well-posed probability distribution).

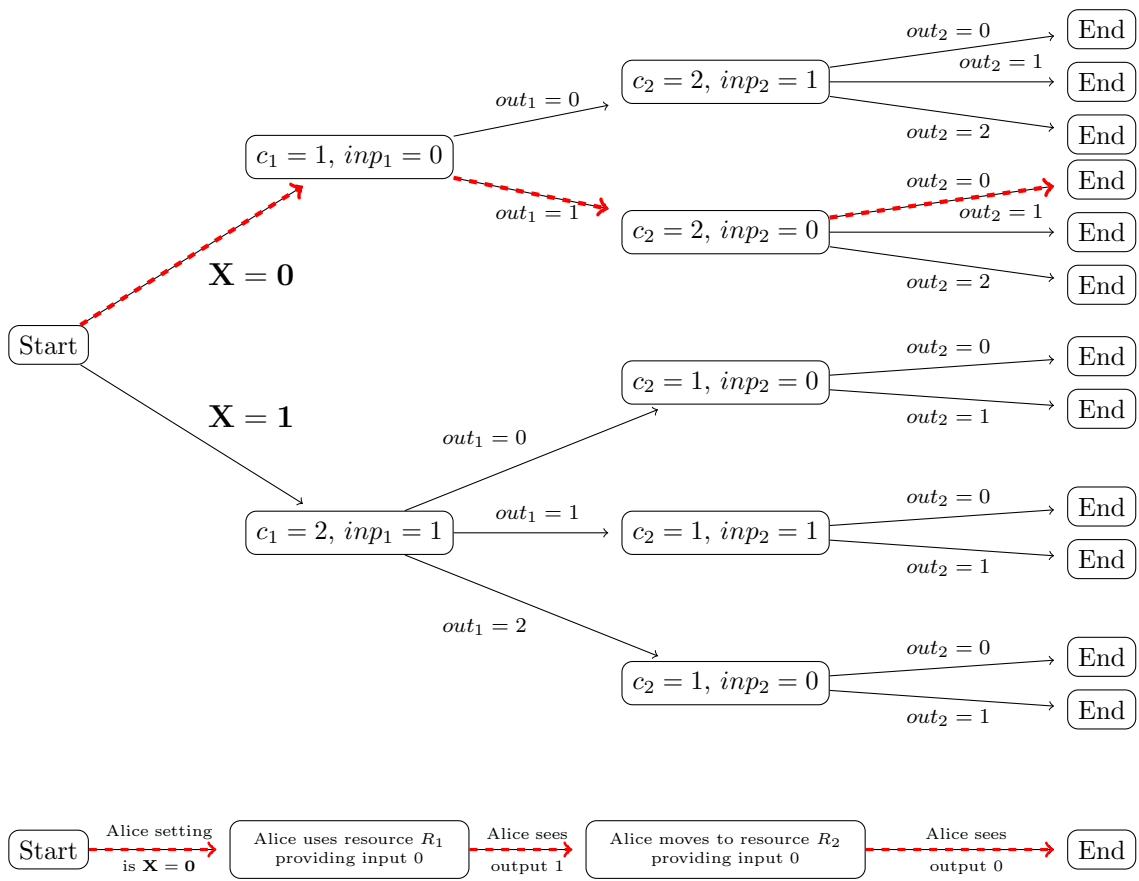
\begin{figure}[t]
\centering
\tikzstyle{bag} = [circle,draw]
\tikzstyle{square} = [rectangle,draw]
\tikzstyle{level 1}=[level distance=4cm, sibling distance=5cm]
\tikzstyle{level 2}=[level distance=5cm, sibling distance=2cm]
\tikzstyle{level 3}=[level distance=5cm, sibling distance=7mm]
\begin{tikzpicture}[grow=right,->,shorten >=5]
\node[rectangle,rounded corners,draw](Start) {\footnotesize Start}
	child {
	node[rectangle,rounded corners,draw,align=center] {\footnotesize $c_1=2$, $inp_1 =1$}     
		child {
           node[rectangle,rounded corners,draw] {\footnotesize $c_2=1$, $inp_2 =0$}  
           	child {
           	node[rectangle,rounded corners,draw] {\footnotesize End}  
			edge from parent 
			node[below]  {\scriptsize $out_2=1$}
			}
           	child {
           	node[rectangle,rounded corners,draw] {\footnotesize End}  
			edge from parent 
			node[above]  {\scriptsize $out_2=0$}
			}
		edge from parent 
		node[below left]  {\scriptsize $out_1=2$}
		}
        	child {
        	node[rectangle,rounded corners,draw] {\footnotesize $c_2=1$, $inp_2 =1$}   
           	child {
           	node[rectangle,rounded corners,draw] {\footnotesize End}  
			edge from parent 
			node[below]  {\scriptsize $out_2=1$}
			}
           	child {
           	node[rectangle,rounded corners,draw] {\footnotesize End}  
			edge from parent 
			node[above]  {\scriptsize $out_2=0$}
			}
      	edge from parent 
        	node[above]  {\scriptsize $out_1=1$}
        	}
		child { 
           node[rectangle,rounded corners,draw] {\footnotesize $c_2=1$, $inp_2 =0$}  
           	child {
           	node[rectangle,rounded corners,draw] {\footnotesize End}  
			edge from parent 
			node[below]  {\scriptsize $out_2=1$}
			}
           	child {
           	node[rectangle,rounded corners,draw] {\footnotesize End}  
			edge from parent 
			node[above]  {\scriptsize $out_2=0$}
			}
		edge from parent 
		node[above left]  {\scriptsize $out_1=0$}
		}
	edge from parent 
	node[above right]  {$\mathbf X=\mathbf 1$}
    	}
	child {
	node[rectangle,rounded corners,draw,align=center](path1) {\footnotesize $c_1=1$, $inp_1 =0$}   
		child {
           node[rectangle,rounded corners,draw](path2) {\footnotesize $c_2=2$, $inp_2 =0$}
           	child {
           	node[rectangle,rounded corners,draw] {\footnotesize End}  
			edge from parent 
			node[below]  {\scriptsize $out_2=2$}
			}
           	child {
           	node[rectangle,rounded corners,draw] {\footnotesize End}  
			edge from parent 
			node[above right]  {\scriptsize $out_2=1$}
			}
           	child {
           	node[rectangle,rounded corners,draw](path3) {\footnotesize End}  
			edge from parent 
			node[above]  {\scriptsize $out_2=0$}
			}
		edge from parent 
		node[below]  {\scriptsize$out_1=1$}
		}
        	child {
        	node[rectangle,rounded corners,draw] {\footnotesize $c_2=2$, $inp_2 =1$}    	          
           	child {
           	node[rectangle,rounded corners,draw] {\footnotesize End}  
			edge from parent 
			node[below]  {\scriptsize $out_2=2$}
			}
           	child {
           	node[rectangle,rounded corners,draw] {\footnotesize End}  
			edge from parent 
			node[above right]  {\scriptsize $out_2=1$}
			}
           	child {
           	node[rectangle,rounded corners,draw] {\footnotesize End}  
			edge from parent 
			node[above]  {\scriptsize $out_2=0$}
			}
        	edge from parent 
        	node[above]  {\scriptsize $out_1=0$}
        	}
	edge from parent 
	node[below right]  {$\mathbf X=\mathbf 0$}
    	}

    ;
    
\draw[dashed, ultra thick,color=red] (Start) -- (path1);
\draw[dashed, ultra thick,color=red] (path1) -- (path2);
\draw[dashed, ultra thick,color=red] (path2) -- (path3);
\end{tikzpicture}

\vspace{1cm}

\begin{tikzpicture}[grow=right,->,shorten >=5]
\node[rectangle,rounded corners,draw](Start) {\footnotesize Start}
	child {
	node[rectangle,rounded corners,draw,align=center](path1) {$\tiny \begin{array}{c}\textnormal{Alice uses resource }R_1\\\textnormal{providing input }0\end{array} $}   
		child {
           node[rectangle,rounded corners,draw](path2) {$\tiny \begin{array}{c}\textnormal{Alice moves to resource }R_2\\\textnormal{providing input }0\end{array} $}   
           	child {
           	node[rectangle,rounded corners,draw](path3) {\footnotesize End}  
			edge from parent 
			node[above]  {\tiny Alice sees}
			node[below]  {\tiny output $0$}
			}
		edge from parent 
		node[above]  {\tiny Alice sees}
		node[below]  {\tiny output $1$}
		}
	edge from parent 
	node[below]  {\tiny is $\mathbf X= \mathbf 0$}
	node[above, align=center] {\tiny Alice setting}
    	}

    ;
    
\draw[dashed, very thick, color=red] (Start) -- (path1);
\draw[dashed, very thick, color=red] (path1) -- (path2);
\draw[dashed, very thick, color=red] (path2) -- (path3);
\end{tikzpicture}
 \caption{\textbf{A decision tree.} The following example is for Alice 1 in the 3-party, 2-resource scenario of Figure \ref{f:3wayexample} where Alice 1 shares resource $R_1$ with Alice 2 (Bob) and Alice 3 (Charlie), and shares resource $R_2$ with Charlie only. Her setting $\mathbf X$ can take one of two values $\{\mathbf 0, \mathbf 1\}$; note that which resource she consults first depends on this setting. Alice can observe one of two possible outputs $\{0,1\}$ from resource $R_1$ and three possible outputs $\{0,1,2\}$ from resource $A_2$. A sample path is highlighted with dashes; Alice's actions and observations for this sample path are detailed at the bottom of the figure.}\label{f:dtree}
\end{figure}

Given a party Alice $p$ who possesses a share of $m_p \le m$ of the resources in $\mathcal R = \{R_1,...,R_m\}$, let us denote the index set of the resources she has access to as $\mathcal R_p$. (Here it is visually useful to note that $\mathcal R_p$ corresponds to $\textit{rows}$ of the output legend in Figure \ref{f:3wayexample}; in contrast the sets $\mathcal M_k$ defined earlier correspond to \textit{columns}.) Then we define a decision tree as follows, with Figure \ref{f:dtree} providing an illustrative example: 

\medskip

\noindent\textbf{Definition.} A \textit{decision tree} for Alice $p$ is a tree graph, consisting of nodes connected by edges, where all maximal length paths (those starting at the root node and ending at a terminal node) are of the same length, exactly $m_p+1$ edges -- the number of resources shared by party $p$, plus one. Furthermore all edges and nodes except for terminal nodes and the root node are labeled, satisfying the following conditions:
\begin{enumerate}
\item There is exactly one edge leaving the root node for each choice of setting $\mathbf x_p$, which is labeled with this setting choice. (This tells Alice $p$ what to do for each choice of setting $\mathbf X_p$).\label{c:setedge}
\item Every non-root and non-terminal node is labeled with two entries instructing Alice $p$ what to do. If $i$ is the number of edges downstream from the root node -- this is known as the depth, or \textbf{level}, of the node -- we notate these two values $(c_i,inp_{i})$, where $c_i$ represents the choice of resource to use, and $inp_i$ represents the choice of input to provide to the corresponding resource $R_{c_i}$. The label $c_i$ is never equal to the an earlier $c_j$ label appearing in one of its ancestor nodes -- a resource is only used once. \label{c:usedonce}
\item For every node carrying a $(c_i,inp_{i})$ label (i.e., non-root, non-terminal nodes), the number of edges descending from it is equal to the number of valid outputs from the resource $R_{c_i}$ for party $p$'s output $A^{(k)}_p$. Each emerging edge is labeled with a unique one of these valid outputs, which we notate $out_i$; the subtree descending from this edge represents what the party proceeds to do conditioned on observing this particular output.\label{e:choiceinput}
\end{enumerate}

\medskip

Fig.~\ref{f:dtree} is an example of a decision tree for Alice 1 in the three-party, two-shared-resources experiment of Figure \ref{f:3wayexample}. The decision tree framework can be slightly augmented if we want to have each party $p$ report an overall outcome upon reaching a terminal node, depending on which terminal node is reached: we model this as the ``final outcome'' $\mathbf A_p$ where disjoint subsets of terminal nodes are labeled with different values $\mathbf a_p$. 

Note that in condition \eref{e:choiceinput}, it is possible for a valid output to occur with probability zero; these are still included on the tree just to avoid some cluttering caveats in the formal arguments below. In a similar vein, conditions \eref{c:setedge}-\eref{e:choiceinput} ensure that in every maximal-length path connecting the root node to a terminal node, the sequence $c_1,...,c_{m_p}$ contained in the traversed nodes maps bijectively to $\mathcal R_p$ (every resource is consulted exactly once). The framework above thus assumes that all parties always use every resource they have access to. This assumption makes the construction of the joint distribution in the next section a little more natural, and is not actually restrictive -- we discuss how to account for the possibility of ``unused" resources further below. 

The key observation about the structure of a party $p$'s decision tree, which enables the sound construction of the joint probability distribution, is as follows: given a setting $\mathbf x_p$, a fixed choice of outputs $a^{(k)}_p$ for each resource $R_k$ shared by party $p$ uniquely determines a max-length path through the decision tree. This is can be confirmed visually by following through Fig.~\ref{f:dtree}: If we are told the setting $\mathbf X_1$ is $\mathbf 0$, then the assignment $A^{(1)}_1=1$ and $A^{(2)}_1=0$ uniquely corresponds to the highlighted path in the tree. Any other assignment $A^{(1)}_1=a^{(1)}_1$ and $A^{(2)}_1=a^{(2)}_1$ corresponds to a different unique branch. Once the branch of the tree is determined, the $inp_i$ labels contained in the nodes along this path specify the inputs that the party must have provided to each resource. Thus each resource input $x^{(k)}_p$ is a function of the string of $a^{(k)}_p$ (as well as the initial setting $\mathbf x_p$), and so could be written $x^{(k)}_p\left(\mathbf x_p, a^{(k)}_{k_1},...,a^{(k)}_{k_{n_k}}\right)$, though we do not use this explicit functional notation below. Indeed, the soundness of the joint probability distribution construction will depend crucially on a further observation that each $x^{(k)}_p$ is determined by only a \textit{proper subset} of the $a^{(k)}_p$: if we work down the initial segment of a path descending to level $i$, this initial segment is determined by $\mathbf X_p$ and $i-1$ choices of $A^{k}_p$, and fixes $i$ choices of $x^{(k)}_p$ that will then be constant independent of the other $A^{(k)}_p$. 

For example: consider a party who possesses a share of five resources $R_1,R_2,R_3,R_4,R_5$, and assume a setting choice $\mathbf X_p = \mathbf x_p$. Consider a fixed string of outputs $a^{(1)}_p, a^{(2)}_p, a^{(3)}_p, a^{(4)}_p, a^{(5)}_p$. Then it is true that the only way that Alice $p$ ends up having observed this specific choice of $a^{(1)}_p, a^{(2)}_p, a^{(3)}_p, a^{(4)}_p, a^{(5)}_p$ is to have traversed a specific unique path through the decision tree, in which she observed these outputs in a certain order and provided specific resource inputs along the way. Suppose that on this path, we have $c_1=3, c_2=2, c_3=5$, so that Alice $p$ must have initially consulted resource $R_3$, then $R_2$, then $R_5$ to be consistent with observing the given output string. Then for other strings of outputs $A^{(1)}_p, a^{(2)}_p, a^{(3)}_p, A^{(4)}_p, a^{(5)}_p$ with any different values of $A^{(1)}_p$ and $A^{(4)}_p$, the corresponding path on the decision tree will have a same initial segment, and map to the same resource input choices of $x^{(2)}_p$, $x^{(3)}_p$, and $x^{(5)}_p$ (and indeed one additional $x^{(k)}_p$ will be determined by $inp_4$, where $k=c_4$). Notice that alternate choices of $A^{(2)}_p, A^{(3)}_p, A^{(5)}_p$ do not necessarily determine $X^{(2)}_p$, $X^{(3)}_p$, and $X^{(5)}_p$: it could be on the decision tree that if $A^{(2)}_p$ is equal to a different $a'^{(2)}_p$, then $c_3$ equals (say) 4 instead of 5 so that $R_4$ is the third resource used; then, we would have instead all strings of the form $A^{(1)}_p, a'^{(2)}_p, a^{(3)}_p, a^{(4)}_p,A^{(5)}_p$ determine the same $x^{(2)}_p, x^{(3)}_p, x^{(4)}_p$ (and one additional $x^{(k)}_p$ determined by $inp_4$ and $c_4$) for all choices of  $A^{(1)}_p$ and $A^{(5)}_p$.

Before moving to the construction of the joint probability distribution, let us briefly return to the question of modeling situations where a party might not use a resource, or may decide to use it only conditionally on seeing certain outputs from other resources. Within the above framework, we can model this a couple of different ways. One option is to introduce an input choice $\perp$ intended to mean ``unused:'' if party $q$ supplies the input $\perp$ for $X_q$, the resource then with probability 1 returns for $A_q$ a ``no output recorded'' result which we also denote as $\perp$; the distribution for the non-$q$ parties $R^{\perp_q \perp_q}(\cdots|\cdots)$ is then just their marginal \eref{e:defreduced}. Another option that avoids the introduction of an extra input choice is to collect all unused resources and put them at the end of the decision tree with an arbitrary dummy choices of input provided, where all outputs lead to the same subtree -- the output is essentially ignored, as the party behaves the same way no matter the output value.

We also note that local probabilistic choices can be encompassed by our framework: if, for instance, Alice $p$ at some point decides to flip a fair coin and condition her input to a later resource based on the coin result, we can model this coin flip as a one-party, single-choice-of-input resource $R_k(A_p|X_p)$ satisfying $R_k(A_p|X_p=\textnormal{the one input})=1/2$. As the degenerate input plays no role, we omit it and represent such resources as $R_k(A_p)$. Multiple parties can also share such input-free resources, which will look like (for example) $R_k(A_2,A_3)$. Such resources correspond to \textit{shared local randomness} and have an operational interpretation as a random process whose output is distributed to the parties prior to the beginning of the experiment. As is well known in the literature, it is convenient to model such choices by combining all of them into a single classical random resource that is shared by all parties; this can be encompassed in our framework and we will return to it more formally later.

We now move on to building the joint distribution from the decision trees of each party, and showing that it is consistent/well-defined (i.e., resulting in a normalized probability distribution).

\subsection{Determining the joint distribution $\mathbb P(\vec A_1, ..., \vec A_m|\mathbf X_1, ..., \mathbf X_n)$. Soundness of the method.}\label{s:probrule}

Given a candidate probability $\mathbb P(\vec a_1, ..., \vec a_m|\mathbf x_1, ..., \mathbf x_n)$ for a fixed choice of $\vec a_1, ..., \vec a_m$ and $\mathbf x_1, ..., \mathbf x_n$, we assign a value between 0 and 1 according to the following method: For each party $p\in \{1,..., n\}$, locate the unique branch (maximal-length path) from their decision tree determined by $\mathbf x_p$ and that party's $a^{(k)}_p$ outputs extracted from among the $\vec a_1, ..., \vec a_m$. Then note the resource inputs $x^{(k)}_j$ that are determined by the $inp_i$ along these paths, and set
\begin{equation}\label{e:probrule}
\fl\mathbb P(\vec a_1, ..., \vec a_m|\mathbf x_1, ..., \mathbf x_n)= \prod_{k=1}^m R_k(\vec a_k|\vec x_k) = \prod_{k=1}^m R_k(a^{(k)}_{k_1}, ...,a^{(k)}_{k_{n_k}}|x^{(k)}_{k_1}, ...,x^{(k)}_{k_{n_k}}) 
\end{equation}
where we have written out $\vec a_k$ as $a^{(k)}_{k_1}, ...,a^{(k)}_{k_{n_k}}$ on the right. The product form of \eref{e:probrule} reflects the intuitive notion that the different resources are indeed different and so operate independently of each other; this eventually underpins the derivation of nontrivial constraints in paradigms such as LOSR-GMNL \cite{coiteux:2021}. The sense in which the resources ``operate independently" is not quite the same as independence of random variables/events with the attendant standard factorization rule $\mathbb P(S \cap T) = \mathbb P(S)\mathbb (T)$: a more relevant (though signaling) analogy would be a scenario of two telephones whose inner workings are completely separate (so ``independent") but one can take what one hears from one telephone (output) and repeat it into the other (as input). Thus in \eref{e:probrule}, while the conditional distributions factor, for a party $p$ an input $x^{(k)}_{p}$ to one resource $R_k$ can depend on (be a function of) an output $a^{(k')}_{p}$ from another resource $R_{k'}$, with the form of the dependence dictated by party $p$'s decision tree. We remark that this notion of independence of resources is important in related but different approaches such as the study of network nonlocality \cite{gisin:2020}.

As an example to illustrate how \eref{e:probrule} is computed, consider the three-party scenario of Figure \ref{f:3wayexample}. Here we have \begin{eqnarray}\label{e:jointexample}
\fl \mathbb P(\vec a_1,  \vec a_2|\mathbf x_1, \mathbf x_2, \mathbf x_3) &=& \mathbb P(\overbrace{a^{(1)}_{1},a^{(1)}_{2},a^{(1)}_{3}}^{\vec a_1}, \overbrace{a^{(2)}_{1},a^{(2)}_{3}}^{\vec{a}_2}|\mathbf x_1, \mathbf x_2, \mathbf x_3)\nonumber\\
&=& R_1(a^{(1)}_{1}, a^{(1)}_2,a^{(1)}_{3}|x^{(1)}_{1}, x^{(1)}_2,x^{(1)}_{3}) R_2(a^{(2)}_{1}, a^{(2)}_3|x^{(2)}_{1}, x^{(2)}_3)\nonumber\\
&=& R_1(a_{1}, a_2,a_{3}|x_{1}, x_2,x_{3}) R_2(a_{1}, a_3|x_{1}, x_3),
\end{eqnarray}
where we remove the $(k)$ superscripts in the last line as they are redundant within an $R_k(\cdots|\cdots)$ expression. If party 1's decision tree is as in Figure \ref{f:dtree}, and if on the left side of \eref{e:jointexample} we have $\mathbf x_1$,  $a^{(1)}_{1}$, and $a^{(2)}_{1}$ as $\mathbf 0$, $1$, and $0$ respectively, then this corresponds to the highlighted path in the figure. The $inp_i$ along this path then determine the corresponding $x^{(1)}_{1}$ and $x^{(2)}_{1}$ values that will appear on the right side of \eref{e:jointexample}; specifically, we then obtain $R_1(1, a_2,a_{3}|0, x_2,x_{3}) R_2(0, a_3|0, x_3)$. To complete the computation of the probability, one would then consult decision trees for the second and third parties to fill in the remaining values.

We claim \eref{e:probrule} yields a valid joint probability distribution for each choice of $\mathbf x_1, ..., \mathbf x_n$. Nonnegativity is immediate: the right side of \eref{e:probrule} is a product of nonnegative terms. Showing that the sum over all values of $\vec a_1, ..., \vec a_m$ is equal to 1 is more involved, and depends critically on the fact that the $R_k$ resources are nonsignaling.

To further motivate our derivations below, let us discuss the potential \textit{failure} of normalization if the resources are signaling. Figure 1 in \cite{bancal:2013} leads to such a failure: here, there are two parties sharing two signaling resources that they access in opposite order resulting in a sort of ``grandfather paradox'' inconsistency as described in further detail in that paper.  Mathematically, if we try to apply our formula \eref{e:probrule} to this example, the right hand side will always take the form $R_1(a_{1}, \beta |\delta, b_2)R_2(\alpha, b_2|a_1,\gamma)$ with the multiple appearances of $a_1$ and $b_2$ resulting from Alice consulting $R_1$ first and using her output as input to $R_2$, while Bob consults $R_2$ first and uses his output as input to $R_1$. Then with the signaling properties of the $R_1$ and $R_2$ distributions as described in \cite{bancal:2013}, at least one of the resources $R_1$ and $R_2$ assigns zero probability for all choices of $a_1$ and $b_2$ independently of the other entries $\alpha,\beta,\gamma,\delta$, so all probabilities assigned by \eref{e:probrule} are zero and thus cannot sum to one. We do not encounter such problems when the nonsignaling condition \eref{e:nosig} is satisfied by the resources.

The proof of normalization, while not immediate, is also not exceedingly involved. However, applying it directly to the general equation \eref{e:probrule} requires some unwieldy notation that can obfuscate what is going on. Hence we first illustrate the key idea with the three-party, two-resource example of Figure \ref{f:3wayexample}. Summing \eref{e:probrule} over all outputs will yield
\begin{equation}\label{e:a1a2a3example}
\fl \sum_{\vec a_1, \vec a_2}\mathbb P(\vec a_1,  \vec a_2|\mathbf x_1, \mathbf x_2, \mathbf x_3) = \sum_{a^{(1)}_1,a^{(1)}_2, a^{(1)}_3,a^{(2)}_1,a^{(2)}_3} R_1(a_{1}, a_2,a_{3}|x_{1}, x_2,x_{3}) R_2(a_{1}, a_3|x_{1}, x_3),
\end{equation}
where we can re-write the right side above in a little more readable fashion, using a standard Alice-Bob-Charlie renaming, as
\begin{equation}\label{e:abcrename}
\sum_{a^{(1)},b^{(1)}, c^{(1)},a^{(2)},c^{(2)}}R_1(abc|xyz) R_2(ac|xz).
\end{equation}
Now, certain $x^{(k)}$, $y^{(k)}$ and $z^{(k)}$ values can depend on $\vec a_1$ and $\vec a_2$ and thus can vary as the sum is performed. However, the input to a party's \textit{first} used resource depends only on their setting $\mathbf x_p$. Let us suppose that for the given choices of $\mathbf x_1,\mathbf x_2,\mathbf x_3$, Alice's and Bob's first steps in their decision trees are to consult resource $R_1$, whereas Charlie consults the other resource $R_2$. Then it is only $z^{(1)}$ and $x^{(2)}$ that can vary in \eref{e:abcrename}. In particular, $z^{(2)}$ is constant in the sum which will allow us to pull a term out as follows. First, re-write \eref{e:abcrename} as
\begin{eqnarray*}
&&\sum_{c^{(2)}}\quad\sum_{a^{(1)},b^{(1)},c^{(1)},a^{(2)}} R_1(abc|xyz) R_2(ac|xz)\\
&=&\sum_{c^{(2)}:R_2(c|z)>0}\quad\sum_{a^{(1)},b^{(1)},c^{(1)},a^{(2)}} R_1(abc|xyz) R_2(ac|xz),
\end{eqnarray*}
where the restriction of the outer sum is valid because given any choice of $c^{(2)}$ for which $R_2(c|z)=0$ holds, $R_2(ac|xz)$ will equal zero as well, and hence the corresponding term in the summand is zero. Then, by \eref{e:condsetout} we can make the substitution
\begin{eqnarray}
R_2(ac|xz)
= R_2(c|z)R_2(a|xz,c)\label{e:cpthenns}
\end{eqnarray} 
and pull out $R_2(c|z)$ to write
\begin{equation}\label{e:firstpullout}
=\sum_{c^{(2)}:R_2(c|z)>0}R_2(c|z)\sum_{a^{(1)},b^{(1)},c^{(1)},a^{(2)}} R_1(abc|xyz) R_2(a|xz,c).
\end{equation}
We remark that this step may fail in the absence of the no-signaling assumption; one would be attempting to pull out $R_2(c|xz)$ instead of just $R_2(c|z)$, and Alice's input $x^{(2)}$ to $R_2$ might not be independent of her output $a^{(1)}$ from $R_1$ (which she consulted first). Continuing on, we can enlist the fact that $x^{(1)}$ and $y^{(1)}$ similarly do not vary in the sum, so that the process can be repeated on the inner sum to rewrite \eref{e:firstpullout} as 
\begin{equation}\label{e:pullout}
\fl \sum_{c^{(2)}:R_2(c|z)>0}R_2(c|z)\sum_{a^{(1)},b^{(1)}:R_1(ab|xy)>0} R_1(ab|xy) \sum_{ c^{(1)},a^{(2)}}R_1(c|xyz,ab)R_2( a|xz,c).
\end{equation}
We have pulled out the probabilities corresponding to the first resource each party consults. 

Now looking at $R_1(c|xyz,ab)$ in the innermost sum, the inputs $x^{(1)}$ and $y^{(1)}$ are fixed, but Charlie's input $z^{(1)}$ can depend on his output $c^{(2)}$ from $R_2$ which he consulted earlier on his decision tree. However, for each choice of $c^{(2)}$ in the outermost sum, $z^{(1)}$ \textit{is} fixed; and for each fixed choice of $a^{(1)}$ and $b^{(1)}$ in the middle sums, $R_1(c|xyz,ab)$ will be a single probability distribution for which we are summing over all outputs $c^{(1)}$ in the innermost sum. We can make a parallel argument for $R_2( a|xz,c)$. So for each fixed choice of $a^{(1)}, b^{(1)},c^{(2)}$ the inner sum is
\begin{equation*}
\sum_{ c^{(1)},a^{(2)}}R_1(c|xyz,ab)R_2( a|xz,c) = \sum_{ c^{(1)}}R_1(c|xyz,ab)\sum_{a^{(2)}}R_2( a|xz,c) = 1.
\end{equation*}
Then \eref{e:pullout} reduces to just the outer sums, for which we have  
\begin{equation*}
\sum_{c^{(2)}:R_2(c|z)>0}R_2(c|z)\sum_{a^{(1)}b^{(1)}:R_1(ab|xy)>0} R_1(ab|xy) = 1.
\end{equation*}

The above example contains the essence of the proof of normalization. For scenarios involving decision trees of depth 3 or more the process of replacing \eref{e:abcrename} with \eref{e:pullout} must be applied iteratively to the inner sum in \eref{e:pullout}: pulling out probabilities corresponding to the second consulted resource on a decision tree, then the third, etc. There is a slight notational complication because the choice of \textit{which} resource is consulted next may change depending on the values fixed by outer sums (i.e., the outputs of the previously consulted resource). For completeness we present the proof of the general case in \ref{s:norm}.

\section{Properties of the induced distribution}\label{s:netproperties}

The distribution defined by \eref{e:probrule} satisfies a number of important properties that we describe here.

\subsection{Nonsignaling of the induced distribution}\label{s:nosig}

We can demonstrate that the distribution $\mathbb P$ defined in \eref{e:probrule} is nonsignaling, in the following sense illustrated for our 3-party example.
Recall that for this example $\mathbb P$ is given by
\begin{equation*}
\mathbb P(\vec a_1,  \vec a_2|\mathbf x_1, \mathbf x_2, \mathbf x_3)
=R_1(a_{1}, a_2,a_{3}|x_{1}, x_2,x_{3}) R_2(a_{1}, a_3|x_{1}, x_3).
\end{equation*}
Now re-ordering the outputs in the front of $\mathbb P(\cdots|\cdots)$ so that they are grouped by party instead of by resource, we can re-write the probability as 
\begin{equation*}
\mathbb P(\underbrace{a^{(1)}_{1},a^{(2)}_{1}}_{\mathbf a_1},\underbrace{a^{(1)}_{2}}_{\mathbf a_2},\underbrace{a^{(1)}_{3},a^{(2)}_{3}}_{\mathbf a_3}|\mathbf x_1, \mathbf x_2, \mathbf x_3).
\end{equation*}
Then, the assertion is that $\mathbb P(\mathbf A_1,\mathbf A_2,\mathbf A_3|\mathbf X_1,\mathbf X_2,\mathbf X_3)$ satisfies the no-signaling condition \eref{e:nosig}: $\sum_{\mathbf a_1} \mathbb P(\mathbf a_1,\mathbf a_2,\mathbf a_3|\mathbf x_1,\mathbf x_2,\mathbf x_3) = \sum_{\mathbf a_1} \mathbb P(\mathbf a_1,\mathbf a_2,\mathbf a_3|\mathbf x'_1,\mathbf x_2,\mathbf x_3)$ for all fixed choices of the non-summed over variables, and the corresponding equalities hold for the parallel expressions with $\sum_{\mathbf a_2}$ and $\sum_{\mathbf a_3}$. Here, we are taking the ``final outcome" of party $p$ -- discussed after the definition of decision trees in Section \ref{s:decisiontree} -- to be the complete transcript of all resource outputs recorded by that party. If party $p$ instead bins together some of these transcripts to report a final outcome as some non-injective function of the complete transcript of resource outputs, the nonsignaling property of the distribution of these binned final outcomes will follow from the nonsignaling property of the distribution of complete transcripts. 

To prove the nonsignaling property in the general $n$-party, $m$-resource setting, recall $\mathcal R_p \subseteq\{1,...,m\}$ denotes the subset of $k$ indices that correspond to resources $R_k$ that are shared by party $p$; these correspond to rows in the output legend of Figure \ref{f:3wayexample}. Then we want to show that for each fixed choice of $p$,
\begin{equation}\label{e:showthat}
\fl\sum_{a^{(k)}_p:k\in \mathcal R_p} \mathbb P(\vec a_{1}, .... , \vec a_{m}| \mathbf x_1,...,\mathbf x_p, ... , \mathbf x_n)=\sum_{a^{(k)}_p:k\in \mathcal R_p} \mathbb P(\vec a_1, .... , \vec a_m| \mathbf x_1,...,\mathbf x'_p, ... , \mathbf x_n)
\end{equation}
for all fixed choices of the non-summed-over variables and settings. We prove this as follows for $p=1$ (the proof applies without loss of generality to the other parties). First, let us deal with a trivial case: suppose that for some resource $k$ shared by party 1, we have $R_k(a_{k_2},...,a_{k_{n_k}}|x_{k_2},...,x_{k_{n_k}})=0$; that is, the marginal probability of the other parties' outputs is zero. Enlisting \eref{e:defreduced}, this implies that for any choice of $x_1$, 
\begin{equation*}
\fl 0 = R_k(a_{k_2},...,a_{k_{n_k}}|x_1,x_{k_2},...,x_{k_{n_k}})=\sum_{a_1} R_k(a_1,a_{k_2},...,a_{k_{n_k}}|x_1,x_{k_2},...,x_{k_{n_k}}) \end{equation*}
and so $R_k(a_1,a_{k_2},...,a_{k_{n_k}}|x_1,x_{k_2},...,x_{k_{n_k}})$ must equal zero for all choices of $a_1$. This implies both sides of the equality in \eref{e:showthat} are zero, after re-expressing $\mathbb P(\cdots|\cdots)$ terms as products of $R(\cdots|\cdots)$ terms according to \eref{e:probrule}. So let us now assume that all $R_k(a_{k_2},...,a_{k_{n_k}}|x_{k_2},...,x_{k_{n_k}})$ are positive. Now using \eref{e:probrule} to write the left side of \eref{e:showthat} as a product of $R(\cdots|\cdots)$ terms, factoring out those not shared by party 1, and applying \eref{e:condsetout}, we can write 
{\footnotesize
\begin{equation*}
\fl\sum_{a^{(k)}_1:k\in \mathcal R_1} \mathbb P(\vec a_1, .... , \vec a_m| \mathbf x_1, ... , \mathbf x_n)=\sum_{a^{(k)}_1:k\in \mathcal R_1}\prod_{k=1}^m R_k(a_{k_1},...,a_{k_{n_k}}|x_{k_1},...,x_{k_{n_k}})
\end{equation*}
\begin{equation*}
\fl=\prod_{k\notin \mathcal R_1}R_k(\cdots|\cdots)\sum_{a^{(k)}_1:k\in \mathcal R_1}\prod_{k\in \mathcal R_1}R_k(\underbrace{a_{k_1}}_{a^{(k)}_1},a_{k_2},...,a_{k_{n_k}}|\underbrace{x_{k_1}}_{x^{(k)}_1},x_{k_2},...,x_{k_{n_k}})
\end{equation*}
\begin{equation*}
\fl=\prod_{k\notin \mathcal R_1}R_k(\cdots|\cdots)\sum_{a^{(k)}_1:k\in \mathcal R_1}\prod_{k\in \mathcal R_1} R_k(a_{k_2},...,a_{k_{n_k}}|x_{k_2},...,x_{k_{n_k}})R_k(a_1|x_1,x_{k_2},...,x_{k_{n_k}},a_{k_2},...,a_{k_{n_k}})
\end{equation*}
\begin{equation*}
\fl=\prod_{k\notin \mathcal R_1}R_k(\cdots|\cdots)\prod_{k\in \mathcal R_1} R_k(a_{k_2},...,a_{k_{n_k}}|x_{k_2},...,x_{k_{n_k}})\sum_{a^{(k)}_1:k\in \mathcal R_1}\prod_{k\in \mathcal R_1} R_k(a_1|x_1,x_{k_2},...,x_{k_{n_k}},a_{k_2},...,a_{k_{n_k}})
\end{equation*}
\begin{equation}
\fl=\prod_{k\notin \mathcal R_1}R_k(\cdots|\cdots)\prod_{k\in \mathcal R_1} R_k(a_{k_2},...,a_{k_{n_k}}|x_{k_2},...,x_{k_{n_k}})\label{e:globalnosig}
\end{equation}
}
\noindent where the last equality holds because the sum in the penultimate line evaluates to one -- a result that can be obtained by noting that this sum is a quantity of the form \eref{e:recursiveform} in \ref{s:norm}, and thus equals 1 by the arguments presented there.\footnote{We can alternatively obtain equality to 1 as a consequence of the normalization of joint probability distributions defined by \eref{e:probrule}: for a fixed choice of $(x_{k_2},...,x_{k_{n_k}})$ and $(a_{k_2},...,a_{k_{n_k}})$, $R_k(a_1|x_1,x_{k_2},...,x_{k_{n_k}},a_{k_2},...,a_{k_{n_k}}) = R_k^{a_{k_2},...,a_{k_{n_k}},x_{k_2},...,x_{k_{n_k}}}(a_1|x_1)$ can be viewed as a one-party nonsignaling resource; then, the product of these reduced $R_k^{a_{k_2},...,a_{k_{n_k}},x_{k_2},...,x_{k_{n_k}}}$ resources appearing in the penultimate line of \eref{e:globalnosig} is the expression for the probability distribution of a one-party network of nonsignaling resources accessed according to party 1's decision tree; summing over all outputs then yields one as a consequence of normalization as proved in Section \ref{s:probrule}.} Now the final expression in \eref{e:globalnosig} has no variables belonging to party 1; all $R_k(a_{k_2},...,a_{k_{n_k}}|x_{k_2},...,x_{k_{n_k}})$ terms depend solely on the settings $\mathbf x_i$ and decision trees of the the other parties. Since the same expression can be reached if we apply these manipulations to the right side of \eref{e:showthat}, the equality holds.

\subsection{Shared local randomness and local deterministic distributions}

Consider an $n$-party paradigm in which arbitrary shared local randomness is allowed: the parties are allowed to consult $n$-party no-input resources $R(a_1,...,a_n)$ of arbitrary distribution. However, restrictions are imposed on the type of nonlocal resources with inputs that can be consulted. An important motivator for this paradigm is the LOSR-GMNL definition of \cite{coiteux:2021}, where global local shared randomness is considered a free resource always available to all $n$ parties, and it is the networks where nonlocal nonsignaling resources (with inputs) are restricted to subsets of two parties that are considered (only) bipartite nonlocal -- or for a more generalized hierarchical notion of LOSR-GMNL \cite{coiteux:2021a}, networks allowing nonlocal nonsignaling resources shared among subsets of at most $n-1$ parties are classified as \textit{not} genuinely $n$-partite nonlocal. 

If we study the class of behaviors satisfying such a paradigm, this class is equivalent to the following: convex mixtures of behaviors induced by networks comprising only extremal nonlocal nonsignaling resources satisfying whatever restrictions were previously imposed on the nonlocal resources. Here, \textit{extremal} resources refers to resources that are extremal in the polytope of nonsignaling resources as defined and discussed in for example \cite{barrett:2005,pironio:2011}; by saying \textit{nonlocal} we specify that these extremal behaviors are not the local deterministic (classical) ones. The equivalence is useful because it means a linear constraint on behaviors induced by networks of extremal nonlocal nonsignaling resources will automatically translate to a linear constraint on the whole class of behaviors by convexity; this technique was used in for example \cite{bierhorst:2021}.

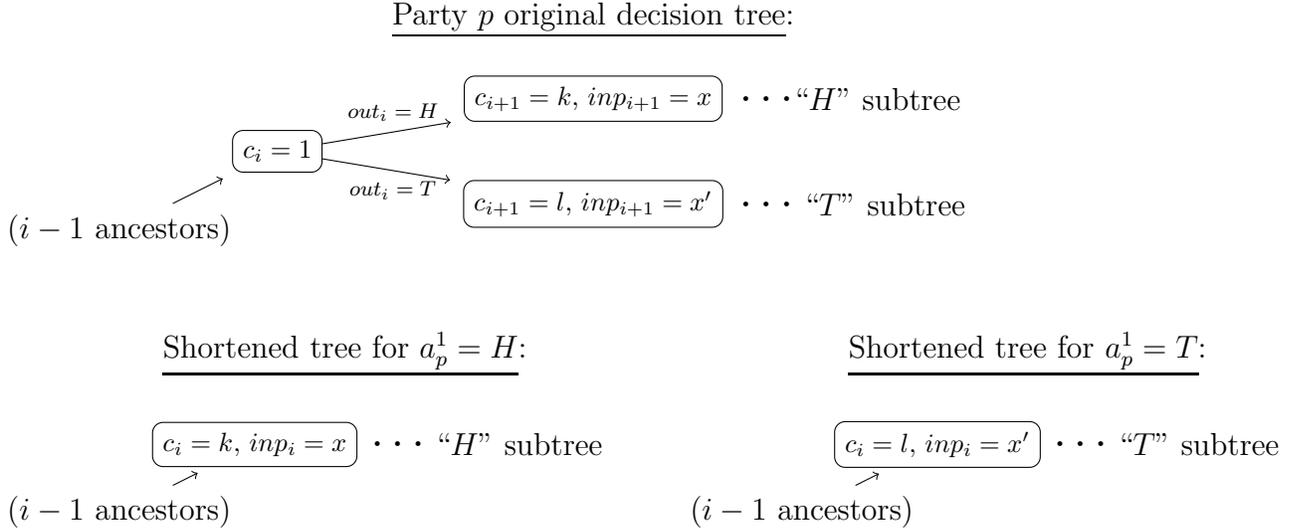
\begin{figure}
\tikzstyle{level 1}=[level distance=3cm, sibling distance=3cm]
\tikzstyle{level 2}=[level distance=6cm, sibling distance=2cm]

\begin{tikzpicture}[grow=right,->,shorten >=5,scale=.7]
\node at (9,4)[align=center]{\underline{Party $p$ original decision tree}:};
\node(Start) {($i-1$ ancestors)}
	child {
	node{}      
	edge from parent[draw=none]}
	child {
	node[rectangle,rounded corners,draw,align=center](path1) {\footnotesize $c_i=1$}  
		child {
           node[rectangle,rounded corners,draw](path2) {\footnotesize $c_{i+1}=l$, $inp_{i+1} =x'$}  node[right=1.8cm]{{\Large$\cdots$} ``$T$'' subtree}
		edge from parent 
		node[below]  {\scriptsize$out_i=T$}
		}
        	child {
        	node[rectangle,rounded corners,draw] {\footnotesize $c_{i+1}=k$, $inp_{i+1} =x$}    	 node[right=1.8cm]{{\Large$\cdots$}``$H$'' subtree}          
        	edge from parent 
        	node[above]  {\scriptsize $out_i=H$}
        	}
    	}

    ;

\end{tikzpicture}

\begin{tikzpicture}[grow=right,->,shorten >=5,scale=.6]
\node at (5,3.5)[align=center]{\underline{Shortened tree for $a^{1}_p=H$}:};
\node(Start) {($i-1$ ancestors)}
	child {
	node{} 
	edge from parent[draw=none]}
	child {
           node[rectangle,rounded corners,draw](path2) {\footnotesize $c_{i}=k$, $inp_{i} =x$}  node[right=1.4cm]{{\Large$\cdots$} ``$H$'' subtree}
    	}
    ;
\end{tikzpicture}
\hspace{.7cm}
\begin{tikzpicture}[grow=right,->,shorten >=5,scale=.6]
\node at (5,3.5)[align=center]{\underline{Shortened tree for $a^{1}_p=T$}:};
\node(Start) {($i-1$ ancestors)}
	child {
	node{}      
	edge from parent[draw=none]}
	child {
           node[rectangle,rounded corners,draw](path2) {\footnotesize $c_{i}=l$, $inp_{i} =x'$}  node[right=1.4cm]{{\Large$\cdots$} ``$T$'' subtree}
    	}
    ;
\end{tikzpicture}
\vspace{-1.5cm}
\caption{\textbf{Removing local randomness from decision trees.} If $R_1$ is a no-input resource, it can be thought of as shared local randomness. Such resources can be removed from decision trees as follows: for party $p$'s original decision tree (top), locate every appearance of the resource (nodes with $c_i=1$), then create a new decision tree for each possible choice of output of the resource (here, there are two outputs) by excising the step where $R_1$ is consulted, bypassing it as though the given choice of output had been observed. Above, we create a new ``$H$'' decision tree by replacing every instance of the top subtree with the shortened below-left subtree, or a new ``$T$'' decision tree by replacing with the shortened below-right subtree.}\label{f:replacingSLR}
\end{figure}

Let us prove the equivalence. The first key property is that shared local randomness resources like $R(a_1,a_3)$ or $R(a_1,...,a_n)$ can always be factored out of the distribution entirely, in the following sense: if $R_1$ is such a resource, we can write
\begin{eqnarray}
\mathbb P(\vec a_1, \vec a_2, ... ,\vec a_m|\mathbf x_1, ... , \mathbf x_n) &=& R_1(\vec a_1) \prod_{k=2}^m R_k(\vec a_k|, \vec x_k)\nonumber\\
&=& R(\vec a_1)\mathbb P_{\vec a_1} (\vec a_2, ... ,\vec a_m|\mathbf x_1, ... , \mathbf x_n)\label{e:factorout}
\end{eqnarray}
where $\mathbb P_{\vec a_1}$ is the distribution that obtains if all parties modify their decision trees as follows: remove all consultations of the resource $R_1$; where such consultations previously occurred, instead proceed directly to the subsequent subtree that followed in the original tree when the output corresponding to $\vec a_1$ was observed. Figure \ref{f:replacingSLR} provides an illustration of this excision/bypassing procedure. Observe that the inputs $\vec x_k$ of the other resources will be the same whether $\mathbb P$ and $\mathbb P_{\vec a_1}$ are expanded according to \eref{e:probrule}, which is why \eref{e:factorout} holds.

A key aspect of \eref{e:factorout} is the following operational interpretation: that of a scenario where the random process $R_1$ is sampled prior to the experiment and the output $\vec a_1$ is distributed to all parties; then when the experiment is run, they proceed with the $\vec a_1$-indexed decision tree. Indeed, if there are multiple local random resources $R_1$ through $R_t$, they can all be pulled out front as
\begin{eqnarray*}
\mathbb P(\vec \mathbf A|\vec \mathbf X) &=& \prod_{k=1}^t R_k(a_1,...,a_n) \prod_{k=t+1}^m R_k(\vec a_k|, \vec x_k)\\
&=& \prod_{k=1}^t R_k(\vec a_k) \mathbb P_{\vec a_1, ..., \vec a_t}(\vec a_{t+1}, ..., \vec a_m),
\end{eqnarray*}
and $\prod_{k=1}^t R_k(\vec a_k)$ can be interpreted as a single combined shared resource $R_{\vec k}$ with the distribution $R_{\vec t}(\vec a_1, ..., \vec a_t) = \prod_{k=1}^t R_k(\vec a_k)$. Hence any behavior in the class $\mathbb P(\vec \mathbf A|\vec \mathbf X)$ is equivalent to a convex mixture of behaviors induced by networks of only resources $R$ with inputs.

A further reduction can be performed. As mentioned earlier, the set of all nonsignaling resource $R(a_1,...,a_n|x_1,...,x_n)$ for a fixed number of parties, inputs, and outputs, comprises a \textit{polytope}, as it is the set of behaviors satisfying linear equalities \eref{e:nosig} along with the linear equalities and inequalities that define valid probability distributions. As such, this polytope will have a certain number $N$ of extreme points $R^{\scriptsize\textnormal{ext}}_i(a_1,...,a_n|x_1,...,x_n)$, $i \in \{1,..., N\}$, for which a general $R(a_1,...,a_n|x_1,...,x_n)$ can be written as a convex combination:
\begin{equation}\label{e:extreme}
R(a_1,...,a_n|x_1,...,x_n) = \sum_i p(i)R^{\scriptsize\textnormal{ext}}_i(a_1,...,a_n|x_1,...,x_n),
\end{equation}
where $p(i)$ is a probability distribution over the values of $i$. Employing such an expression for $R_1(\vec a_1| \vec x_1)$, we can write
\begin{eqnarray}
\mathbb P(\vec a_1, \vec a_2, ... ,\vec a_m|\mathbf x_1, ... , \mathbf x_n) &=& R_1(\vec a_1|\vec x_1) \prod_{k=2}^m R_k(\vec a_k|, \vec x_k)\nonumber\\
&=& \left[\sum_{i}p(i)R^{\scriptsize\textnormal{ext}}_i(\vec a_1|\vec x_1)\right] \prod_{k=2}^m R_k(\vec a_k|, \vec x_k)\nonumber\\
&=& \sum_{i}p(i)\left[R^{\scriptsize\textnormal{ext}}_i(\vec a_1|\vec x_1)\prod_{k=2}^m R_k(\vec a_k|, \vec x_k\right] \label{e:replaceextremal}
\end{eqnarray}
and the term in brackets in \eref{e:replaceextremal} is equal to $\mathbb P_i(\vec a_1, \vec a_2, ... ,\vec a_m|\mathbf x_1, ... , \mathbf x_n)$, which we define to be the distribution induced when each party uses their original decision with the single change of replacing consultations of $R_1$ with consultations of $R^{\scriptsize\textnormal{ext}}_i$. Hence $\mathbb P$ is equal to the convex mixture $\sum_i p(i) \mathbb P_i$. This process can be repeated for all $R_k$ so that $\mathbb P$ is a convex mixture of distributions each induced by extremal-only resources.

As a final simplification, consider local deterministic distributions. These distributions are extremal in the nonsignaling polytope, but they do not exhibit nonlocal behavior: they are classical where each party's output is a deterministic function of their local input. That is, there is a function $f$ mapping each input $x_i$ to a fixed value in the range of $A_i$ for which
\begin{equation*}
R(a_1,...,a_n|x_1,...,x_n) = \prod_{i=1}^n \delta_{a_i,f(x_i)}
\end{equation*}
where the Kronecker delta function $\delta_{a_i,f(x_i)}$ maps to one if $a_i=f(x_i)$ and zero otherwise. Such a resource can be removed from decision trees with no meaningful consequences for the induced distribution $\mathbb P$ by merely bypassing all steps where it is consulted: there is only one relevant edge descending from a consultation of $R$, the one labeled with the probability-one output $f(x_i)$. Specifically, for a local deterministic $R_1$ we can write $\mathbb P (\vec a_1,\vec a_2, ... ,\vec a_m|\mathbf x_1,...,\mathbf x_n) = \mathbb P' (\vec a_2, ... ,\vec a_m|\mathbf x_1,...,\mathbf x_n) \prod_{i=1}^n \delta_{a_i,f(x_i)}$ where $\mathbb P'$ is the distribution resulting from the shortened decision trees; $\mathbb P$ has no meaningful characteristics not already possessed by $\mathbb P'$.

\subsection{A discussion of causality}

The inflation technique \cite{wolfe:2019} is an important tool for deriving constraints on behaviors that can arise in multiparty networks. It applies to all theories that are \textit{causal}, which includes quantum mechanics, the scenario of wired nonsignaling boxes studied in this paper, and even more general probabilistic theories that would allow generalized analogs of entangled measurements on the nonsignaling resources. It is accepted in \cite{coiteux:2021,coiteux:2021a} that the theory of wired nonsignaling boxes is causal. However, as discussed in the introduction, some references cited regarding this question \cite{janotta:2012,chiribella:2011,chiribella:2014} are somewhat abstract/general and focused on other questions, not explicitly addressing the matter in regards to wired nonsignaling boxes. It is thus useful to spend a few paragraphs discussing how the results derived in earlier sections imply the causality of wired nonsignaling boxes.

A definition of a causal theory amenable to the inflation technique can be found in Section IIB of \cite{coiteux:2021a}: a theory is causal if it satisfies the conditions of Definition 1 therein along with ``device replicability." We paraphrase the proffered definition of causality loosely as follows: consider a theory with multipartite resources (for us, the $R_k$ resources), parties who measure them (Alice 1, Alice 2,...), and rules that determine the probabilities observed by the parties given the resources they measure (for us, the decision trees of Section \ref{s:defandconsist} and the induced probability rule \eref{e:probrule}). Then the theory is \textit{causal} if it satisfies the following conditions: first, given a subset of parties along with all the resources they share -- some of which may be additionally shared with parties outside the subset -- the subset parties' marginal distribution will be the same regardless of how the measured resources are connected to (or disconnected from, or re-connected to) parties outside the subset: for example, in Figure \ref{f:3wayexample}, if we look at just the two parties Alice and Bob, their marginal distribution should be the same even if resource $R_1$ is connected to one Charlie-type party while resource $R_2$ is measured by a different Charlie-type party (where the different Charlies, in turn, are perhaps measuring different $R_1$ and $R_2$ resources connected to other Alice and Bob-type parties, etc.). A ``Charlie-type" party is a measuring party with the same decision tree. 

The second condition, somewhat implicit in the wording of the first given above, is that it makes sense to speak of multiple copies of $R_1$ and $R_2$ resources: the theory should allow the devices to be replicated, and if a network is sound in the sense that each resource is always connected to an appropriate party and vice versa (for example, a resource like $R_2$ in Fig.~\ref{f:3wayexample} is always connected to a party like Alice and a party like Charlie; Charlie is always connected to a resource of form $R_1$ and $R_2$), then the theory provides a sound probability distribution for the parties of the network. (This is key for the mechanics of the inflation technique: a network of interest is ``inflated" to a larger network locally isomorphic to the original one; straightforward constraints on the probability distribution of the larger network reveal subtler insights about the smaller network -- for this to work, it is necessary that the  larger network \textit{have} a probability distribution.) 

The third condition is independence of distributions among parties with no common resources: if two parties measure no common resources, then their joint distribution should factor: $\mathbb P(\mathbf A\mathbf B|\mathbf X \mathbf Y) = \mathbb P(\mathbf A|\mathbf X ) \mathbb P(\mathbf B|\mathbf Y)$ if there is no resource $R$ measured by both Alice and Bob. This condition is also required to hold more generally for two disjoint subsets of parties.

We now discuss how the theory of wired nonsignaling resources satisfies the conditions of causality described above. First, the nonsignaling condition derived in Section \ref{s:nosig} ensures that the marginal distribution of a subset of parties does not depend on how the resources they share are connected to parties outside the subset. Specifically, looking back to \eref{e:globalnosig}, we see that if we remove Alice 1 from consideration, the marginal distribution of the remaining parties is the product of the resource distributions $R(\cdots|\cdots)$ with Alice 1's variables removed. Thus if we have a set of $n'<n$ parties of interest, we can remove the non-$n'$ parties one by one until only the $n'$ parties remain, at which point their marginal distribution is (uniquely) determined as the product of the resources they share, now treated as the marginal resources $R(a_1,...,a_{n'}|x_1,...,x_{n'})$ that remain when the other parties' $a_i$ and $x_i$ variables have been removed. This will be the same expression regardless of how the removed parties were connected or not connected to resources, with the resulting marginal distribution only depending on the decision trees of the $n'$ parties. 

The second condition (device replication) follows from the soundness of the probability formula \eref{e:probrule} for always producing a consistent probability distribution from cascaded measurements of nonsignaling resources $R_k$, as discussed at length and proved in Section \ref{s:probrule} and \ref{s:ap1}. Finally, the third condition (independence) follows from \eref{e:probrule} when we consider that 
\begin{eqnarray*}
&&\prod_{k=1}^m R_k(a_{k_1}, ...,a_{k_{n_k}}|x_{k_1}, ...,x_{k_{n_k}})\\
&=& \prod_{k\in \mathcal S} R_k(a_{k_1}, ...,a_{k_{n_k}}|x_{k_1}, ...,x_{k_{n_k}})\prod_{k\in \mathcal S^C} R_k(a_{k_1}, ...,a_{k_{n_k}}|x_{k_1}, ...,x_{k_{n_k}})\\
\end{eqnarray*} 
and if each party either shares resources only from $\mathcal S$, or only from $\mathcal S^C$, then the factors above will correspond to respective distributions $\mathbb P_\mathcal S$ and $\mathbb P_{\mathcal S^C}$ that would obtain individually from two disjoint networks treated separately, so that $\mathbb P$ factors into the product of $\mathbb P_\mathcal S$ and $\mathbb P_{\mathcal S^C}$.

\section{An application: deriving the Chao-Reichardt inequality \cite{chao:2017} and others from that of Mao \textit{et al.}~\cite{mao:2022}}

We now turn to inequalities witnessing LOSR-GMNL in the three-party setting. Chao and Reichardt \cite{chao:2017} give an early example of a constraint on three-party behaviors induced by wired networks of 2-party-only nonlocal nonsignaling resources, with access to global shared (local) randomness; a linear version of this constraint is given in \cite{bierhorst:2021} where it is derived rigorously. The arguments in \cite{bierhorst:2021,chao:2017} directly work with the nonsignaling resources and do not invoke the inflation technique. Later constraints introduced by \cite{coiteux:2021,coiteux:2021a} and improved upon in \cite{mao:2022,cao:2022} employed the inflation technique and so constrain a more general class of behaviors (allowing for additional features in the bipartite-only networks such as entangled measurements of quantum resources). 

In the context of the previous section, which solidifies the applicability of the inflation technique to wired nonsignaling boxes, it is notable to show how the early inequality of \cite{bierhorst:2021,chao:2017} can be obtained from that of \cite{mao:2022}: this exercise provides an alternate proof of the Chao-Reichardt inequality, and shows it constrains a more general class of theories (i.e., all causal theories as opposed to just wired nonsignaling boxes). Let us stipulate that each of the parties have two settings and outcomes, where the settings $X,Y,Z$ take a value in $\{0,1\}$ and the outcomes $A,B,C$ take a value in $\{-1,+1\}$. (Note in previous sections we denoted $A,B,C,X,Y,Z$ with boldface; we do not do so here to align with the notation of \cite{mao:2022}.) Define
\begin{equation*}
\langle A_x B_y\rangle = \mathbb P(A=B|xy)-\mathbb P(A\ne B|xy)
\end{equation*}
so that the above is equal to the expected value $\mathbb E(AB|xy)$ of the product of the outcomes of $A$ and $B$. Similarly, we can define a three-way expectation
\begin{equation*}
\langle A_xB_yC_z\rangle= \mathbb P(ABC=+1|xyz)-\mathbb P(ABC=-1|xyz).
\end{equation*}
Now the inequality of Mao \textit{et al.}~\cite{mao:2022} is
\begin{equation}\label{e:mao}
\langle A_0B_0\rangle +\langle A_0B_1\rangle+\langle A_1B_0C_1\rangle -\langle A_1B_1C_1\rangle +2\langle A_0C_0\rangle \le 4
\end{equation}
(see expression (3) in this reference), where the above must hold of networks of bipartite-only nonlocal resources but can be violated if the three-way entangled GHZ resource is measured. The 3-party inequality of Chao and Reichardt \cite{chao:2017} is formulated in \cite{bierhorst:2021} (see expressions (14) and (15) therein) as
\begin{equation*}
\fl 4\mathbb P(A\ne C|X=0,Z=0) + P(A\ne B|X=0,Y=0)+P(A\ne B|X=0,Y=1)
\end{equation*}
\begin{equation}\label{e:chao}
\fl+\mathbb P(ABC=-1|X=1,Y=0,Z=1)+\mathbb P(ABC=+1|X=1,Y=1,Z=1)\ge 1 
\end{equation}
which if we re-write in terms of $\langle AB\rangle$ type expressions using conversions of the form $\mathbb P(A=B|xy)=(1+\langle A_x B_y\rangle)/2$ and $\mathbb P(A\ne B|xy)=(1-\langle A_x B_y\rangle)/2$ along with their three-party analogs, and perform some algebra, we get
\begin{equation*}
\langle A_0 B_0\rangle+\langle A_0 B_1\rangle
+\langle A_1 B_0C_1\rangle-\langle A_1 B_1C_1\rangle + 4\langle A_0C_0\rangle \le 6
\end{equation*}
which can be obtained from \eref{e:mao} by adding the trivial algebraic inequality $\langle A_0C_0\rangle\le 1$ twice. \eref{e:mao} is evidently the stronger constraint.

Interestingly, the inequality of Cao \textit{et al.}~\cite{cao:2022} can be derived from \eref{e:mao} as well: If we re-label Bob's outcomes when his setting is 1 by interchanging $+1$ and $-1$, \eref{e:mao} becomes
\begin{equation}\label{e:mao1}
\langle A_0B_0\rangle -\langle A_0B_1\rangle+\langle A_1B_0C_1\rangle +\langle A_1B_1C_1\rangle +2\langle A_0C_0\rangle \le 4.
\end{equation}
Switching the roles of Alice and Charlie in \eref{e:mao1} yields
\begin{equation}\label{e:mao2}
\langle C_0B_0\rangle -\langle C_0B_1\rangle+\langle A_1B_0C_1\rangle +\langle A_1B_1C_1\rangle +2\langle A_0C_0\rangle \le 4,
\end{equation}
and adding \eref{e:mao1} and \eref{e:mao2} together yields
\begin{equation}\label{e:cao}
\fl \langle A_0B_0\rangle +\langle B_0C_0\rangle -\langle A_0B_1\rangle-\langle B_1C_0\rangle+4\langle A_0C_0\rangle +2\langle A_1B_0C_1\rangle +2\langle A_1B_1C_1\rangle \le 8
\end{equation}
which is the 3-party inequality (1) in \cite{cao:2022}. Thus all known (3,2,2) inequalities (3-party, 2-outcome, 2-setting) can be derived from that of Mao \textit{et al.}~\cite{mao:2022}. Note that since both \eref{e:mao1} and \eref{e:mao2} require genuine tripartite nonlocality to violate, their sum \eref{e:cao} should not necessarily be considered a weaker witness of LOSR-GMNL when compared to \eref{e:mao}.

Reference \cite{cao:2022} contains another inequality (S14) in the supplementary material which in the 3-party case is not a (3,2,2) inequality (Bob has a third setting) but it can be obtained from \eref{e:mao} nonetheless. The three party version of (S14) is
\begin{eqnarray}
\fl \frac{1-\langle C_1\rangle}{2}\left(\langle A_0B_0\rangle_{C=-1,Z=1}-\langle A_0B_1\rangle_{C=-1,Z=1}+\langle A_1B_0\rangle_{C=-1,Z=1}+\langle A_1B_1\rangle_{C=-1,Z=1}\right)\nonumber\\
\fl+\frac{1+\langle C_1\rangle}{2}\left(\langle A_0B_0\rangle_{C=+1,Z=1}+\langle A_0B_1\rangle_{C=+1,Z=1}+\langle A_1B_0\rangle_{C=+1,Z=1}-\langle A_1B_1\rangle_{C=+1,Z=1}\right)\nonumber\\
+\langle A_0B_2\rangle +\langle B_2 C_0\rangle \le 6,\label{e:cao2}
\end{eqnarray}
where $\langle A_xB_y\rangle_{C=c,Z=1}$ is the expectation conditioned on $C=c,Z=1$. It turns out that the first two lines of \eref{e:cao2} are equivalent to the first four terms of \eref{e:mao}, and the fifth term of \eref{e:mao} can be replaced using the algebraic inequality $\langle A_0 C_0\rangle \ge \langle A_0B_2\rangle +\langle B_2 C_0\rangle-1$ (this inequality is used in \cite{coiteux:2021a,cao:2022,wolfe:2019} and can be confirmed by writing out all the probabilities), leading to \eref{e:cao2}. Details of the derivation are given in \ref{s:maocao}.

The only other three party inequality currently known to witness LOSR-GMNL is (1) in \cite{coiteux:2021a}, which was tested in \cite{huang:2022}. Like \eref{e:cao2}, this inequality has a third setting for Bob, and while (1) of \cite{coiteux:2021a} admits a linear form \cite{patra:2024} it does not appear to be directly derivable from \eref{e:mao}.

\section{Conclusion}

We have shown the consistency of probability distributions induced by wired nonsignaling resources, shown that such distributions are themselves nonsignaling, and discussed other properties such as causality and the ability to factor out classical random resources and ignore local deterministic distributions while restricting attention to extremal nonsignaling resources. This study was motivated in part by new definitions of Genuine Multipartite Nonlocality (the ``LOSR-GMNL" definition of \cite{coiteux:2021}), and we closed with an example showing how most inequalities witnessing tripartite GMNL can be derived from that of \cite{mao:2022}. Going forward, the framework developed in this paper will provide a useful foundation for rigorously proving future results about wired nonsignaling resources; this will be useful in studying the gap between this scenario and more general scenarios permitting entangled measurements -- notably, the inflation technique constrains all causal theories and so cannot directly target this gap. The results here are also relevant to other paradigms, such as for example networks of quantum-achievable nonsignaling resources measured in cascaded fashion as studied in the proposed definition of genuine network nonlocality given in \cite{supic:2022}. Future work may also explore generalizations to encompass resources that admit some restricted form of signaling, such as in models that utilize underlying one-way signaling resources to replicate quantum nonlocal behaviors \cite{bancal:2013,bancal:2012}, to see under what weaker conditions a consistent joint distribution as in \eref{e:probrule} may still be guaranteed.

\medskip

\noindent \textbf{Acknowledgments.} This work was partially supported by NSF Award No.~2210399 and AFOSR Award No.~FA9550-20-1-0067. 

\appendix

\section{Proofs of nonsignaling properties}\label{s:ap1}

Here we write out a couple of the longer equation sequences proving claims in Section \ref{s:properties}. First, we write out how \eref{e:nosigsubcomp} is a consequence of \eref{e:nosig}. The three party version of this argument appears in Section IIIA of \cite{barrett:2005}  which we merely iterate more times to obtain the following:
\begin{eqnarray*}
&&\sum_{\vec a_q} R(\vec a_p, \vec a_q | \vec x_p, \vec x_q)\\
&=&\sum_{a_{p+1},...,a_{n}}  R(\vec a_p, a_{p+1}, ...,  a_n| \vec x_p, x_{p+1}, ... ,  x_n)\\
&=& \sum_{a_{p+1},...,a_{n-1}} \left(\sum_{a_n} R(\vec a_p, a_{p+1}, ..., a_{n-1}, a_n| \vec x_p, x_{p+1},x_{p+2}, ... , x_{n-1}, x_n)\right)\\
&=& \sum_{a_{p+1},...,a_{n-1}} \left(\sum_{a_n} R(\vec a_p, a_{p+1}, ..., a_{n-1}, a_n| \vec x_p, x_{p+1},x_{p+2}, ... , x_{n-1}, x'_n)\right)\\
&=& \sum_{a_{p+1},...,a_{n-2},a_n} \left(\sum_{a_{n-1}} R(\vec a_p, a_{p+1}, ..., a_{n-1}, a_n| \vec x_p, x_{p+1},x_{p+2}, ..., x_{n-1}, x'_n)\right)\\
&=& \sum_{a_{p+1},...,a_{n-2},a_n} \left(\sum_{a_{n-1}} R(\vec a_p, a_{p+1}, ..., a_{n-1}, a_n| \vec x_p, x_{p+1},x_{p+2}, ... , x'_{n-1}, x'_n)\right)\\
&\vdots&\\
&=&  \sum_{a_{p+2},...,a_n} \left(\sum_{a_{p+1}} R(\vec a_p, a_{p+1}, ..., a_{n-1}, a_n| \vec x_p, x_{p+1},x'_{p+2}, ... , x'_{n-1}, x'_n)\right)\\
&=& \sum_{a_{p+2},...,a_n} \left(\sum_{a_{p+1}} R(\vec a_p, a_{p+1}, ..., a_{n-1}, a_n| \vec x_p, x'_{p+1},x'_{p+2}, ... , x'_{n-1}, x'_n)\right)\\
&=&\sum_{\vec a_q} R(\vec a_p, \vec a_q | \vec x_p, \vec x'_q).
\end{eqnarray*}
The steps above alternate between re-arranging order of summation, and then applying \eref{e:nosig} to the inner sum within parentheses. The above proof does not depend on the ordering of the parties; choosing $\vec x_p$ as an initial string just makes it easier to notate. The condition thus applies to any two complementary sets of parties.

We now show that our definition of $R(\vec a_p|\vec x_p)$ in \eref{e:defreduced} is consistent with what we would find for this quantity from a direct manipulating conditional probabilities, if we model the inputs $X_i$ as random variables with a probability distribution (which along with the specification of $R(\vec A_n |\vec X_n)$ induces a full joint distribution of $\vec A_n$ and $\vec X_n$). Assuming $R(\vec x_p)>0$ -- if not, $R(\vec a_p|\vec x_p)$ is undefined -- we can write, for any choice $\vec a_p$,
\begin{eqnarray}
R(\vec a_p|\vec x_p) &=& R(\vec a_p, \vec x_p)/R(\vec x_p)\nonumber\\
&=&\left[\sum_{\vec x_q:R(\vec x_p, \vec x_q)>0}R(\vec a_p,\vec x_p, \vec x_q)\right]/R(\vec x_p)\nonumber\\
&=&\left[\sum_{\vec x_q:R(\vec x_p, \vec x_q)>0}R(\vec a_p|\vec x_p, \vec x_q)R(\vec x_p, \vec x_q)\right]/R(\vec x_p)\nonumber\\
&=&\left[\sum_{\vec x_q:R(\vec x_p, \vec x_q)>0}R(\vec a_p|\vec x_p, \vec x^*_q)R(\vec x_p, \vec x_q)\right]/R(\vec x_p)\nonumber\\
&=&\left[R(\vec a_p|\vec x_p, \vec x^*_q)\sum_{\vec x_q:R(\vec x_p, \vec x_q)>0}R(\vec x_p, \vec x_q)\right]/R(\vec x_p)\nonumber\\
&=&\left[R(\vec a_p|\vec x_p, \vec x^*_q)R(\vec x_p)\right]/R(\vec x_p)\nonumber\\
&=&R(\vec a_p|\vec x_p, \vec x^*_q)\nonumber,
\end{eqnarray}
where $\vec x^*_q$ is a fixed choice of values for $\vec X_q$, which allows for pulling the term $R(\vec a_p|\vec x_p, \vec x^*_q)$ out of the sum over $\vec x_q$ after previously replacing each (varying) choice of $\vec x_q$ in $R(\vec a_p|\vec x_p, \vec x_q)$ with this fixed $\vec x^*_q$ by invoking \eref{e:compcomp}. Since $\vec x^*_q$ above can be any value of $\vec x_q$ for which $R(\vec x_p, \vec x_q)>0$, defining $R(\vec a_p|\vec x_p)$ in \eref{e:defreduced} as $R(\vec a_p|\vec x_p,\vec x_q)$ for any fixed choice of $\vec x_q$ is sensible.

\section{Normalization in the general setting}\label{s:norm}

To prove normalization of \eref{e:probrule} in the general case, we rely on the following arithmetic construction. Suppose a quantity $Q$ can be written as
\begin{equation}\label{e:recursive}
Q=\sum_i \xi_i f(i), \quad \textnormal{with } \sum_i \xi_i = 1
\end{equation}
where each $f(i)$ is a number which may vary with $i$. If $f(i)$ happens to equal 1 for all choices of $i$, then $Q=1$, but we do not initially assume this is the case. We assume instead that each $f(i)$ can be written in a form parallel to \eref{e:recursive}:
\begin{equation*}
f(i) = \sum_j\eta^i_{j} g^i(j), \quad \textnormal{with } \sum_j\eta^i_{j} = 1
\end{equation*}
where $g^i(j)$ is a number that may vary with $j$, and then we say that $Q$ satisfies the \textit{recursive sum-to-1 property} if the process can always be repeated such that each new nested functional term can be written in the form of \eref{e:recursive}, while assuming that this process eventually terminates in a final expression of the form \eref{e:recursive} where the functional term \textit{does} equal one uniformly (i.e., not varying with the summed index). Then with a little thought, we see that the original quantity $Q$ must equal one as follows: each bottom-level sum, for which the functional term is uniformly one, will itself equal one; then move back up one level where the functional terms are now known to be 1, and the next-higher-level sum will equal one as well; continuing to recursively work back up to the original quantity $Q$ level by level, we find $Q=1$.

For our problem, we show that for a fixed setting choice $\mathbf{x}_1,...,\mathbf{x}_n$, the sum of all $\mathbb P (\vec a_1, ..., \vec a_m | \mathbf x_1,...,\mathbf x_n)$ terms, which is the quantity
\begin{equation}\label{e:induct1}
\sum_{\vec a_1, ...,\vec a_m} \prod_{k=1}^m R_k(a^{(k)}_{k_1}, ...,a^{(k)}_{k_{n_k}}|x^{(k)}_{k_1}, ...,x^{(k)}_{k_{n_k}}),
\end{equation}
satisfies this recursive sum-to-1 property, and thus equals 1. The idea is to successively perform the manipulation that took us from \eref{e:abcrename} to \eref{e:firstpullout} as we work down a decision tree until eventually what is left in the inner sum is a sum over a single variable that is equal to one.

As a first step, pick a party $p$ -- for ease of notation, let us say it is Alice 1 -- and consult this party's decision tree to find the first resource they consult after being provided setting $\mathbf x_1$; denote this $R_{t}$ and let $x^{(t)}_{t_1}=x^{(t)}_1$ denote the input specified by $inp_1$. Now in \eref{e:induct1}, limit the sum over $a^{(t)}_{1}$ to precisely those values for which $R_{t}(a_{1}|x_{1})>0$, which does not change the value of the sum as $R_{t}(a_{1}|x_{1})=0$ implies that the term $R_{t}(a_{t_1},a_{t_2},..., a_{t_{n_t}}|x_{t_1},x_{t_2},..., x_{t_{n_t}})$ appearing in the summand will be zero as well. (Recall that ${t_1},{t_2},..., {t_{n_t}}$ denotes the indices of the subset of parties sharing resource $R_t$, so for this resource $a^{(t)}_{t_1}=a^{(t)}_1$ and $x^{(t)}_{t_1}=x^{(t)}_1$.) Now for each value of $a^{(t)}_1$ for which $R_{t}(a_1|x_1)>0$ we can write
\begin{equation*}
\fl R_{t}(a_{1},a_{t_2},..., a_{t_{n_t}}|x_{1},x_{t_2},..., x_{t_{n_t}})=R_{t}(a_{t_2},..., a_{t_{n_t}}|x_{1},x_{t_2},..., x_{t_{n_t}}, a_{1})R_{t}(a_{1}|x_{1})
\end{equation*}
via the same manipulation that was performed in \eref{e:cpthenns}.  Then since $x_{1}$ is determined by $\mathbf x_1$ alone, we can pull $R_{t}(a_{1}|x_{1})$ out of the sum and re-write \eref{e:induct1} as follows:
\begin{equation}\label{e:after1}
\fl \underbrace{\sum_{a^{(t)}_{1}:R_{t}(a_{1}|x_{1})>0} R_{t}(a_{1}|x_{1})}_{\sum_i \xi_i}\underbrace{\sum_{\{\vec a_1, ...,\vec a_m\}\backslash a^{(t)}_{1}} R_1(\cdots|\cdots)R_2(\cdots|\cdots)\cdots R_k(\cdots|\cdots)}_{f(i)},
\end{equation}
where the $R(\cdots|\cdots)$ terms of the inner sum are as in \eref{e:induct1} except for $R_{t}(\cdots|\cdots)$ which now equals $R_{t}(a_{t_2},..., a_{t_{n_t}}|x_{1},x_{t_2},..., x_{t_{n_t}}, a_{1})$. Now, following \eref{e:defreduced} the terms $R_{t}(a_{1}|x_{1})$ constitute a probability distribution and so will sum to one, justifying the labeling with $\sum_i \xi_i$ above, so the above expression is a $Q$-type quantity as in \eref{e:recursive}. The value of the inner sum can vary with the choice of index of the outer sum, as is allowed for the $f(i)$ terms in \eref{e:recursive}.

To perform the inductive step of the argument, we show that that the terms labeled $f(i)$ in \eref{e:after1} satisfy certain general conditions, and that these conditions (alone) ensure that each $f(i)$ term can always be re-written in a form $\sum_j \eta_j g(j)$ with $\sum_j \eta_j=1$ \textit{such that} that the same general conditions will hold for each $g(j)$; thus the process will always be repeatable, and as a final step we will see that it terminates in an expression uniformly equaling one. The general conditions are motivated by the idea that we will pull out terms from the inner sum one by one, with each pull-out corresponding to taking a single step down a party's decision tree to the next consulted resource. 

Now we lay out the conditions: each term labeled $f(i)$ in \eref{e:after1} (which vary with the outer sum) is an expression of the form
\begin{equation}\label{e:recursiveform}
\sum _{M \subseteq \{\vec a_1,..., \vec a_m\}}R_1(\cdots|\cdots)\cdots R_m(\cdots | \cdots)
\end{equation}
where summing over $M \subseteq \{\vec a_1,..., \vec a_m\}$ is to be understood that that a subcollection of variables of the form $a^{(k)}_j$ are being summed over. If we think of ourselves as working down decision trees, $M$ corresponds to parties' ``pending outputs" from resources that have not yet been consulted. The following properties are satisfied by expression \eref{e:recursiveform}: 

\begin{enumerate}
\item For each $a^{(k)}_p \in M^C$ -- that is, an $a^{(k)}_p$ that is not being summed over -- a fixed choice $a^{(k)}_p $ appears in the conditioner of $R_k(\cdots|\cdots)$, along with a fixed choice of $x^{(k)}_p$, and these do not vary. (These correspond to resources that have already been consulted, having worked part way down a path on a decision tree.)\label{cond1}
\item For each party $p$, if we collect all the fixed values $a^{(k)}_p$ from $M^C$ for this choice of $p$, these determine an initial path in party $p$'s decision tree descending from the overall setting choice $\mathbf x_p$. The fixed $x^{(k)}_p $ appearing inside the summand are consistent with the $inp_i$ on this initial path.\label{cond2}
\item For each $a^{(k)}_j$ in $M$, $a^{(k)}_j $ appears (varying) in the front of the appropriate $R_k(\cdots|\cdots)$ term, and the corresponding $x^{(k)}_j$ will appear in the conditioner. These $x^{(k)}_j$ are not necessarily fixed and may change as the sum over $M$ is performed -- they are determined by the fixed choice of $a$ values from $M^C$ along with the varying-with-the-sum choices of $a$ values from $M$.\label{cond3}
\item We adopt a convention that any term of the form $R_k( \emptyset | \cdots)$ -- that is, with no terms in the front -- equals one. This corresponds to a resource that all parties have already consulted and so the corresponding $a^{(k)}_j$ are all in $M^C$. (Operationally, this should be understood to indicate a resource $R_k$ that has been pulled out of the inner sum completely; however we leave a rump term behind with this notational oddity to help maintain the inductive form \eref{e:recursiveform} through all steps.)\label{cond4}
\end{enumerate}

Now we show that the conditions ensure that \eref{e:recursiveform} can be re-written as $\sum_j \eta_j g(j)$ with $\sum_j \eta_j=1$ and each $g(j)$ also satisfying the conditions. To do so, consult the part of a party $p$'s decision tree that is determined by that party's initial setting $\mathbf x_p$ along with the fixed choices of that party's $a$ values from the collection $M^C$ (if any), which by conditions \eref{cond1} and \eref{cond2} does determine a unique (initial) path for party $p$. Let $i$ be the length of this initial path. Then it will determine a choice of resource $c_i$ and input $inp_i$ to be used at the next step; thus for the resource $R_t(\cdots|\cdots)$ with $t=c_i$, the value of  $x^{(t)}_p$ in the conditioner will be fixed as $inp_i$ for all terms of the sum in \eref{e:recursiveform} (even as the corresponding $a^{(t)}_p\in M$ varies as it is summed over). For ease of notation let us assume that $p=1$, so $R_t(\cdots|\cdots)$ will appear in \eref{e:recursiveform} as
\begin{equation}\label{e:appears}
R_t(a_1, \vec a_q|x_1, \vec x_q, \vec x_r,\vec a_r)=R_t^{\vec x_r,\vec a_r}(a_1, \vec a_q|x_1, \vec x_q)
\end{equation}
where $\vec a_q$ are among the $M$ indices and $\vec a_r$ are among the $M^C$ indices. For values of $a_1$ for which $R_t(a_1|x_1, \vec x_r,\vec a_r)=R_t^{\vec x_r,\vec a_r}(a_1|x_1)$ is nonzero, we can apply \eref{e:condsetout} to $R_t^{\vec x_r,\vec a_r}$ to re-write the above expression as 
\begin{eqnarray*} 
R_t(a_1, \vec a_q|x_1, \vec x_q, \vec x_r,\vec a_r)=R_t(\vec a_q|x_1, \vec x_q, \vec x_r,a_1,\vec a_r)R_t(a_1|x_1, \vec x_r,\vec a_r)
\end{eqnarray*}
where the equality follows from the fact that this sort of conditioning can be performed iteratively as discussed following \eref{e:outcomeconditional}. Now pull out $R_t(a_1|x_1, \vec x_r,\vec a_r)$ to re-write \eref{e:recursiveform} as follows:
\begin{equation}\label{e:recursivestep}
\fl\sum_{a^{(t)}_1:R_t(a_1|x_1, \vec x_r,\vec a_r)>0}R_t(a_1|x_1, \vec x_r,\vec a_r)\sum _{M' =M\backslash a^{(t)}_1}R_1(\cdots|\cdots)\cdots R_m(\cdots | \cdots)
\end{equation}
where the inner summand above differs from the summand in \eref{e:recursiveform} by replacing $R_t(a_1, \vec a_q|x_1, \vec x_q, \vec x_r,\vec a_r)$ with $R_t(\vec a_q|x_1, \vec x_q, \vec x_r,a_1,\vec a_r)$. Now, the first sum in \eref{e:recursivestep} is the $\sum_i \xi_i$ portion of \eref{e:recursive}, where $\sum_i \xi_i=1$ as $R_t(a_1|x_1, \vec x_r,\vec a_r)$ is a probability distribution over $a_1$. And we now can argue that for each fixed choice of $a^{(t)}_1$,
\begin{equation}\label{e:endrecursion}
\sum _{M'=M\backslash a^{(t)}_1}R_1(\cdots|\cdots)\cdots R_m(\cdots | \cdots)
\end{equation}
is an expression of the form \eref{e:recursive} satisfying the general conditions \eref{cond1}-\eref{cond4}, where we have effectively moved $a^{(t)}_1$ from $M$ to $M^C$. To elaborate: as required by \eref{cond1} for $M'^C$, a fixed choice of $a^{(t)}_1$ now appears in the \textit{conditioner} of $R_t(\cdots|\cdots)$, and as noted preceding \eref{e:appears} the choice of $x^{(k)}_1$ will be fixed as well. The initial path for Alice 1 referred to in \eref{cond2} is now one level longer in \eref{e:endrecursion} compared to \eref{e:recursiveform}, while still satisfying the condition. For all other parties, satisfaction of \eref{cond1} and \eref{cond2} in \eref{e:recursiveform} carries over immediately to \eref{e:endrecursion}. Finally, satisfaction of \eref{cond3} carries over directly from \eref{e:recursiveform} to \eref{e:endrecursion} as $M'\subset M$.

Regarding the eventual termination of this process, each round of induction moves an $a$ variable from the front of an $R_k(\cdots|\cdots)$ term to the conditioner; there is a finite number of times this will occur before all terms remaining in \eref{e:endrecursion} are of the form described in condition \eref{cond4} -- and so $M'$ is the empty set -- at which point \eref{e:endrecursion} equals 1, completing the argument.

\section{Obtaining (S14) in \cite{cao:2022} from (1) in \cite{mao:2022}}\label{s:maocao}

In this appendix we explain how the first two lines of \eref{e:cao2} are equivalent to the first four terms of \eref{e:mao}. Thus the expression \eref{e:cao2} is a consequence of \eref{e:mao} when $2\langle A_0 C_0\rangle$ is replaced according to an algebraic inequality described in the main text.

Substituting $(1+\langle C_1\rangle)/2=\mathbb P(C=+1|Z=1)$ and $(1-\langle C_1\rangle)/2=\mathbb P(C=-1|Z=1)$, we rewrite \eref{e:cao2} as 
\begin{eqnarray}
\fl\mathbb P(C=-1|Z=1)\left(\langle A_0B_0\rangle_{C=-1,Z=1}+\langle A_0B_1\rangle_{C=-1,Z=1}-\langle A_1B_0\rangle_{C=-1,Z=1}+\langle A_1B_1\rangle_{C=-1,Z=1}\right)\nonumber\\
\fl+\mathbb P(C=+1|Z=1)\left(\langle A_0B_0\rangle_{C=+1,Z=1}+\langle A_0B_1\rangle_{C=+1,Z=1}+\langle A_1B_0\rangle_{C=+1,Z=1}-\langle A_1B_1\rangle_{C=+1,Z=1}\right)\nonumber\\
+\langle A_0B_2\rangle +\langle B_2 C_0\rangle \le 6.\label{e:cao2a}
\end{eqnarray}
Writing out expectations in a more explicit form
\begin{equation*}
\langle A_xB_y\rangle_{C=c,Z=1}=\mathbb E(AB|X=x,Y=y,Z=1; C=c) 
\end{equation*}
and noting that by no-signaling
\begin{equation*}
\mathbb P(C=c|Z=1)= \mathbb P(C=c|X=x,Y=y,Z=1),
\end{equation*}
we can rewrite the sum of two terms $\mathbb P(C=-1|Z=1)\langle A_0B_0\rangle_{C=-1,Z=1}$ and ${\mathbb P(C=+1|Z=1)\langle A_0B_0\rangle_{C=+1,Z=1}}$ that appear in \eref{e:cao2a} after multiplying out as
\begin{eqnarray*}
&& \mathbb E(AB|X=0,Y=0,Z=1; C=-1)\mathbb P(C=-1|X=0,Y=0,Z=1)\\
&+&\mathbb E(AB|X=0,Y=0,Z=1; C=+1)\mathbb P(C=+1|X=0,Y=0,Z=1)\\
&=& \mathbb E(AB|X=0,Y=0,Z=1)
\end{eqnarray*}
by the law of iterated expectation; $\mathbb E(AB|X=0,Y=0,Z=1)$ is equal to ${\mathbb E(AB|X=0,Y=0) = \langle A_0B_0\rangle}$ in turn by nonsignaling. A similar argument yields $\langle A_0B_1\rangle$ from two other terms of \eref{e:cao2a} after multiplying out. On the other hand, we can write $-\langle A_1B_0\rangle_{C=-1,Z=1}$ as
\begin{equation*}
\fl -\left[\mathbb P(A=B|X=1,Y=0,Z=1; C=-1) - \mathbb P(A\ne B|X=1,Y=0,Z=1;C=-1) \right]
\end{equation*}
which when we multiply by $\mathbb P(C=-1|Z=1)=\mathbb P(C=-1|X=1,Y=0,Z=1)$ becomes
\begin{equation*}
\fl -\mathbb P(A=B,C=-1|X=1,Y=0,Z=1) +\mathbb P(A\ne B,C=-1|X=1,Y=0,Z=1);
\end{equation*}
performing a similar calculation on $\langle A_1B_0\rangle_{C=+1,Z=1}\mathbb P(C=+1|Z=1)$ and adding the result to the expression above, the resulting sum can be equivalently re-written as
\begin{equation*}
\fl \mathbb P(ABC=+1|X=1,Y=0,Z=1) - \mathbb P(ABC=-1|X=1,Y=0,Z=1) = \langle A_1B_0C_1\rangle.
\end{equation*}
Similarly, the final remaining two $\langle A_xB_y\rangle$ type terms in \eref{e:cao2a} are equivalent to $-\langle A_1B_1C_1\rangle$.

\medskip
\begin{center}
{\huge \textbf{References}}
\end{center}
\medskip
\bibliographystyle{unsrt}
\bibliography{metabib}

\end{document}